\documentclass[preprint,amsmath,amssymb,prd,a4]{revtex4}
\usepackage{graphicx}
\usepackage{bm}
\usepackage{txfonts}
\usepackage{hyperref}
\usepackage{bm}
\usepackage{graphicx}
\usepackage{diagbox}

\date{\today}
\newcommand{\be}{\begin{eqnarray}}
	\newcommand{\ee}{\end{eqnarray}}

\newcommand{\bfb}{{\bf b}_{\perp}}

\newcommand{\bfP}{{\bf P}_{\perp}}

\newcommand{\bfp}{{\bf p}_{\perp}}

\newcommand{\Dp}{{\bf \Delta}_{\perp}}

\usepackage{xcolor}
\usepackage{rotating}

\begin{document}
	%
	%
	\title{Deciphering Twist-3 Chiral-Even GPDs in the Light-Front Quark-Diquark Model
}
 \author{Sameer Jain}
	\email{sameerjainofficial@gmail.com}
	\affiliation{Department of Physics, Dr. B. R. Ambedkar National Institute of Technology, Jalandhar, 144008, India}
	\author{Shubham Sharma}
	\email{s.sharma.hep@gmail.com}
	\affiliation{Department of Physics, Dr. B. R. Ambedkar National Institute of Technology, Jalandhar, 144008, India}
	\author{Harleen Dahiya}
	\email{dahiyah@nitj.ac.in}
	\affiliation{Department of Physics, Dr. B. R. Ambedkar National Institute of Technology, Jalandhar, 144008, India}
	
	\date{\today}
	%
\begin{abstract}
	We investigate quantum chromodynamics (QCD) in this study by computing chiral-even generalized parton distributions (GPDs) at twist-$3$ using the light-front quark-diquark model (LFQDM), particularly when the longitudinal momentum transfer is zero. We provide a detailed analysis of the twist-$3$ chiral-even GPD's dependence on the longitudinal momentum fraction ($x$) and the momentum transfer ($t$) by illustrating their behavior through extensive two-dimensional ($2$-D) and three-dimensional ($3$-D) visualizations.  Our investigation also reveals the intricate relationships between these GPDs and other distribution functions (DFs) such as generalized transverse-momentum dependent distributions (GTMDs), transverse momentum-dependent parton distributions (TMDs), and parton distribution functions (PDFs). Our study also includes the connected form factors (FFs) which are crucial in understanding the internal structure of hadrons. Additionally, we provide impact parameter GPD plots to offer insights into the spatial distribution of partons.
	\par
	\vspace{0.1cm}
\end{abstract}
%
\maketitle
\section{Introduction\label{secintro}}
\noindent
The proton is among the first baryon to be discovered experimentally, yet, even after a century, its structure remains elusive \cite{Gao:2017yyd}. The `proton spin crisis' has intrigued researchers for decades, and the origin of proton mass is a highly active field of research, encompassing both experimental and theoretical investigations \cite{Altarelli:1988nr, Anselmino:1994gn, Ball:1995td, Bass:2004xa, Altarelli:1998nb, Bissey:2005kd, Wilczek:2012sb, Ji:1994av, Lorce:2017xzd, Metz:2020vxd, Wang:2019mza}. Experimental facilities such as the Stanford Linear Accelerator Center (SLAC) \cite{E142:1993hql, E143:1998hbs, E154:1997xfa, E155:1999eug, E155:2000qdr, E155:2002iec}, the European Organization for Nuclear Research (CERN) \cite{SpinMuonSMC:1997voo, SpinMuonSMC:1997mkb, SpinMuon:1998eqa, COMPASS:2010wkz, COMPASS:2015mhb, COMPASS:2016jwv}, the Deutsches Elektronen-Synchrotron (DESY) \cite{HERMES:1998cbu, HERMES:2006jyl, HERMES:2011xgd}, and Jefferson Lab (JLab) \cite{CLAS:2003rjt, Kramer:2005qe, CLAS:2006ozz, CLAS:2014qtg, Deur:2014vea} have played crucial roles in aforementioned studies. The usual approach of analyzing the proton's structure includes scattering experiments: deep inelastic scattering (DIS) being a key method \cite{Collins:1981uk, Ji:2004wu, Lai:2010vv}. Using the factorization theorem, the cross-section of DIS is parameterized in terms of parton distribution functions (PDFs), which are quasi-probabilistic distributions providing information about the partons, i.e., quarks or gluons, inside the proton \cite{Chay:2013zya, Gluck:1994uf, Collins:1981uw}. Although at very short distances, i.e., at very high energies, perturbative quantum chromodynamics (QCD) provides significant results by adding leading order (LO), next-to-leading order (NLO), and higher-order corrections \cite{Chen:2024fhj, Bjorken:1968dy}, at low energies, factorization theorems are not as effective \cite{Wang:2024wny}. Moreover, nonperturbative effects become dominant. To gain a clearer understanding of a proton's structure at low energies, it is necessary to consider additional corrections, including target mass corrections and higher twist corrections \cite{Wilson:1969zs, Brandt:1970kg, Christ:1972ms}.
\par
\begin{figure*}
	\centering
	\begin{minipage}[c]{0.98\textwidth}
		\includegraphics[width=17cm]{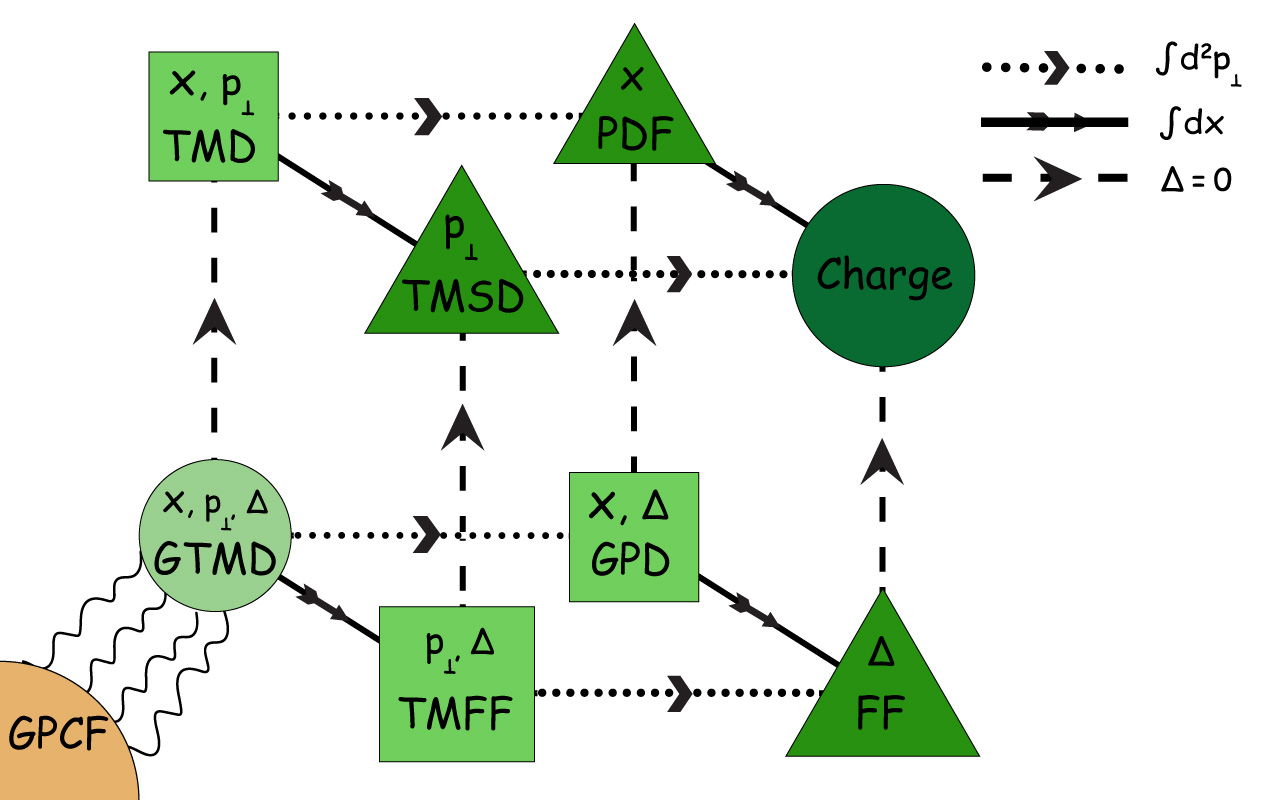}
		\hspace{0.05cm}\\
	\end{minipage}
	\caption{\label{figtree} A illustration of the generalized Parton Correlation Functions (GPCFs) using family trees. Various arrows represent various GTMD limits. The dashed line represents the case of zero momentum transfer, the solid line represents the integration over the longitudinal momentum fraction $x$, and the dotted line shows the integration over the quark's transverse momentum $\bfp$ \cite{Sharma:2023tre}.
	}
\end{figure*}

Information obtained from PDFs is somewhat restricted as it acknowledges only the one-dimensional ($1$-D) distribution of the longitudinal momentum fraction $x$ of the parton. Higher-dimensional distributions such as transverse momentum distributions (TMDs) provide more information concerning the proton's three-dimensional ($3$-D) configuration using kinematic variables such as $x$ and the transverse momentum of the parton $p_{\perp}$ \cite{Collins:1981uk, Ji:2004wu, Cahn:1978se, Konig:1982uk, Chiappetta:1986yg, Collins:1984kg, Sivers:1989cc, Efremov:1992pe, Collins:1992kk, Collins:1993kq, Kotzinian:1994dv, Mulders:1995dh, Boer:1997nt, Boer:1997mf, Boer:1999mm, Bacchetta:1999kz, Brodsky:2002cx, Collins:2002kn, Belitsky:2002sm, Burkardt:2002ks, Pobylitsa:2003ty, Goeke:2005hb, Bacchetta:2006tn, Cherednikov:2007tw, Brodsky:2006hj, Avakian:2007xa, Miller:2007ae, lattice-TMD, Arnold:2008kf, Brodsky:2010vs}. TMDs correspond to phenomena such as semi-inclusive deep inelastic scattering (SIDIS) and Drell-Yan (DY) \cite{Falciano:1986wk, Conway:1989fs, Zhu:2006gx, Arneodo:1986cf, Airapetian:1999tv, Avakian:2003pk, Airapetian:2004tw, Alexakhin:2005iw, Gregor:2005qv, Ageev:2006da, Airapetian:2005jc, Kotzinian:2007uv, Diefenthaler:2005gx, Osipenko:2008rv, Giordano:2009hi, Airapetian:2009jy}.
Another higher-dimensional distribution: the generalized parton distributions (GPDs) are parameterized in the variables $x$ and momentum transfer to the proton $\Dp$. Such distributions correspond to scattering processes such as deeply virtual Compton scattering (DVCS) and deeply virtual meson production (DVMP) \cite{Mueller:1998fv, Diehl03, Ji04, Belitsky05, Boffi07, Ji96, Brodsky06, Radyushkin97, Burkardt00, Diehl02, DC05, Hagler03, Kanazawa14, Rajan16, Ji97, Radyushkin:1996nd, Ji:1998pc, Goeke:2001tz}. Even greater-dimensional distributions, such as generalized transverse momentum distributions (GTMDs), carry the most detailed information about the parton \cite{Lorce13}. The relationship between these distributions, along with some other distributions, can be seen in Fig. \ref{figtree}.
\par
GPDs appear while studying the cross-section of scattering processes such as DVCS and DVMP. They are also known as the non-forward matrix elements of bilocal operators in such processes. Although direct extraction of GPDs from sophisticated experiments such as Zentrum für Elektronen-Und-Speicherringexperiment (ZEUS) \cite{ZEUS:2003pwh, ZEUS:2008hcd}, CEBAF Large Acceptance Spectrometer (CLAS) \cite{CLAS:2007clm, CLAS:2008ahu, Niccolai:2012sq, CLAS:2015bqi, CLAS:2015uuo, CLAS:2018bgk, CLAS:2018ddh, CLAS:2021gwi}, and Common Muon and Proton Apparatus (COMPASS) \cite{Kumericki:2016ehc} is not straightforward, it becomes even more complicated when we consider higher twist corrections \cite{Bertone:2021yyz, Moffat:2023svr}. Refs. \cite{Lorce:2014mxa, Bhoonah:2017olu} suggest that twist-$3$ GPDs can provide vital information about quark's kinetic orbital angular momentum (OAM) and quark spin-orbit interactions. The Fourier transform of GPDs provides the impact parameter dependent parton distribution functions (IPDPDFs), which are functions of $x$ and the impact parameter distance $\bf{b_{\perp}}$ \cite{Burkardt:2002hr}. IPDPDFs offer the most physical picture of the proton as they suggest the position of partons, giving a familiar understanding of the structure of an object.
\par 
One of the most successful and profound theories in the history of physics, offering explanations for physical phenomena with unparalleled accuracy, particularly in the realm of high-energy physics, is quantum field theory (QFT). The remarkable achievements of QFT are best illustrated by QCD, the theory of quark-gluon interactions \cite{Brodsky:1997de}. However, achieving this level of accuracy presents tremendous computational challenges that require sophisticated methods and adjustments that consider various effects. One mathematical approach to simplifying computations in QCD is the use of Anti-de Sitter (AdS)/QCD correspondence. This method, along with Dirac's light-front dynamics, can dramatically simplify calculations and provide more familiarity with the underlying physics \cite{Harindranath:1996hq}. 
The light-front quark-diquark model (LFQDM) serves as a model that uses both of these mathematical formulations along with the assumption that during interactions with a probe, the proton acts as a composite of an active quark participating in the interaction while the remaining quarks form a spectator diquark \cite{Maji:2016yqo, Chakrabarti:2019wjx, Maji:2017bcz}.
In recent years, LFQDM has accomplished many significant results. For example, it has shown very promising results for spin asymmetry, aiding in the study of experiments like Hadron-Elektron Ring Anlage Measurement of Spin (HERMES) and COMPASS \cite{Gurjar:2022rcl}. The flavor combination of the PDF $e(x)$ compares nicely with the CLAS data \cite{Gurjar:2022rcl}. Multiple properties of the proton, such as mechanical radius, shear forces, and pressure distributions, along with structure functions such as gravitational form factors (FFs) and transversity and helicity PDFs, have been calculated using LFQDM \cite{Chakrabarti:2020kdc}. Recent works also include calculations of the transverse structure of the proton in Refs. \cite{Maji:2017bcz, sstwist3, sstwist4}. Twist-$2$ and twist-$4$ GTMDs are discussed in Refs. \cite{Sharma:2023tre, majigtmd}, while twist-$2$ and twist-$4$ GPDs have been calculated in Refs. \cite{Sharma:2023ibp, Maji:2017ill}.
\par
The objective of our work aims to analyze the twist-$3$ GPDs of proton within LFQDM framework. Primarily, the unintegrated quark-quark GPD correlator has been deciphered for twist-$3$ Dirac matrix structure, and we then, through comparison with the parameterization equations, achieve the equations for the twist-$3$ GPDs of the proton. The explicit equations for GPDs have been derived for both possible scenarios of active quark flavor $u$ and $d$ from vector and scalar diquark components, considering a skewness $\xi$ of $0$. The nature of twist-$3$ chiral-even GPDs is illustrated using two-dimensional ($2$-D) and $3$-D graphs, depicting their dependency over quark's longitudinal momentum fraction $x$ and the momentum transfer $t$. To unify the obtained  findings of this study and their connections with other distribution functions, we seek the associated GTMDs, TMDs, and PDF. We have included the analysis of associated GPDs in impact parameter space, obtained through the Fourier transformation of GPDs. Additionally, twist-$3$ FFs have also been discussed considering their significance for the comprehension of proton dynamics.
\par
The article has adopted the following structure: The LFQDM's crucial details, input parameters, and other constants are discussed in Sec. \ref{secmodel}. The twist-$3$ quark-quark GPD correlator features are covered in Sec. \ref{seccor}, along with the pertinent parameterization equations. The explicit equations for twist-$3$ GPDs are shown in Sec. \ref{secresults}. A sequential analysis of the relationships between twist-$3$ chiral-even GPDs and GTMDs along with TMDs is presented in Sec. \ref{secrel1} and \ref{secrel2} respectively. A $2$-D and $3$-D plot-based analysis of GPDs is presented in Sec. \ref{secdiscussion}. This section also covers Fourier-transformed GPD illustrations and twist-$3$ FFs. Finally, a conclusion is presented in Sec. \ref{seccon}.
\section{Light-Front Quark-Diquark Model (LFQDM) \label{secmodel}}
\noindent
Regarding the LFQDM explanation, for an all-encompassing perspective on the probability of running into every possible active quark-spectator combination, the proton's spin-flavor structure is thought to be composed of isoscalar-scalar diquark singlet $|u~ S^0\rangle$, isoscalar-vector diquark $|u~ A^0\rangle$, and isovector-vector diquark $|d~ A^1\rangle$ states from Ref. \cite{Maji:2016yqo}
\begin{equation}
	|P; \Lambda^{N} \rangle = C_S|u~ S^0\rangle^{\Lambda^{N}} + C_V|u~ A^0\rangle^{\Lambda^{N}} + C_{VV}|d~ A^1\rangle^{\Lambda^{N}}. \label{PS_state}
\end{equation}
In the above expression, the nucleon helicity is $\Lambda^{N}$. The spin-wise vector and scalar diquark parts are denoted by $A=V,~VV$ and $S$, respectively. The diquarks respective isospins have been indicated by the superscripts $(0)$ or $(1)$. The coefficients $C_{i}$ of scalar and vector diquark states have been found in Ref. \cite{Maji:2016yqo} and are provided in Table \ref{tab_par}. The valence quark's proportion of longitudinal momentum from the parent proton is $x=p^+/P^+$, where the momentum of quark ($p$) and diquark ($P_X$) are given as
\begin{eqnarray}
	p &&\equiv \bigg(xP^+, p^-,\bfp \bigg)\,,\label{qu} \\
	P_X &&\equiv \bigg((1-x)P^+,P^-_X,-\bfp\bigg), \label{diq}
\end{eqnarray}
for the case when proton carries no transverse momenta.
The expansion of Fock-state for ${J^{z}} =\pm 1/2$ for the scalar $|\nu~ S\rangle^{\Lambda^{N}} $ and vector diquark $|\nu~ A \rangle^{\Lambda^{N}}$ in the case of two particles can be expressed as \cite{Ellis:2008in,Maji:2017bcz,Lepage:1980fj}
\begin{eqnarray}
		|u~ S\rangle^\pm &=&\sum_{\lambda^q}  \int \frac{dx~ d^2\bfp}{2(2\pi)^3\sqrt{x(1-x)}}  \psi^{\pm(\nu)}_{\lambda^q}\left(x,\bfp\right)\bigg|\lambda^{q},\lambda^{S}; xP^+,\bfp\bigg\rangle ,\label{fockSD}\\
  |\nu~ A \rangle^\pm &=&\sum_{\lambda^q} \sum_{\lambda^D} \int \frac{dx~ d^2\bfp}{2(2\pi)^3\sqrt{x(1-x)}} \Bigg[ \psi^{\pm(\nu)}_{\lambda^q \lambda^D }\left(x,\bfp\right)\bigg|\lambda^{q},\lambda^{D}; xP^+,\bfp\bigg\rangle \label{fockVD}.
\end{eqnarray}
The flavor index $\nu=u$ (for the scalar case) and $\nu=u,d$ (for the vector case) are determined using Eq. (\ref{PS_state}). The two particle state is represented by the expression $|\lambda^{q},~\lambda^{Sp};  xP^+,\bfp\rangle$, where the quark helicity is $\lambda^{q}=\pm\frac{1}{2}$ and the spectator diquark helicity is $\lambda^{Sp}$. The scalar diquark's spectator helicity is $\lambda^{Sp}=\lambda^{S}=0$ (singlet), while the vector diquark's spectator helicity is $\lambda^{Sp}=\lambda^{D}=\pm 1,0$ (triplet). Table \ref{tab_LFWF} provides the LFWFs \cite{Maji:2017bcz} for ${J^{z}}=\pm1/2$, taking into account the scalar or vector nature of diquarks.
\begin{table}[h]
	\begin{center}
		\centering
	\begin{tabular}{ |p{1.24cm}|p{1.2cm}|p{0.97cm}|p{0.82cm} p{3.9cm}|p{0.82cm} p{3.9cm}|}
		\hline
		\rule{0pt}{6.5mm}$\hspace{-0.23mm}\rm{Diquark}~$~&$~~~~\lambda^{q}$&$~~\lambda^{Sp}$&\multicolumn{2}{c|}{LFWFs for $J^z=+1/2$} & \multicolumn{2}{c|}{LFWFs for $J^z=-1/2$} \\[1.9 mm] \hline
	
		\rule{0pt}{6.5mm}~~$\hspace{-0.69mm} \rm{Scalar}~$&~$+1/2$&~$~~~0$&$\psi^{+(\nu)}_{+}$&$~=~N_S~ \varphi^{(\nu)}_{1}$&$\psi^{-(\nu)}_{+}$&$~=~N_S \bigg(\frac{p^1-ip^2}{xM}\bigg)~ \varphi^{(\nu)}_{2}$   \\[1.9 mm]
		\rule{0pt}{6.5mm}&~$-1/2$&~$~~~0$&$\psi^{+(\nu)}_{-}$&$~=~-N_S\bigg(\frac{p^1+ip^2}{xM} \bigg)~ \varphi^{(\nu)}_{2}$&$\psi^{-(\nu)}_{-}$&$~=~N_S~ \varphi^{(\nu)}_{1}$~    \\[2.9 mm]
		\hline
		\rule{0pt}{6.5mm}$\hspace{-0.13mm}~\rm{Vector}$&~$+1/2$&$~~+1$&$\psi^{+(\nu)}_{+~+}$&$~=N^{(\nu)}_1 \sqrt{\frac{2}{3}} \bigg(\frac{p^1-ip^2}{xM}\bigg)~  \varphi^{(\nu)}_{2}$&$\psi^{-(\nu)}_{+~+}$&$~=~0$~   \\[1.9 mm] 
		\rule{0pt}{6.5mm}&~$-1/2$&$~~+1$&$\psi^{+(\nu)}_{-~+}$&$~=~N^{(\nu)}_1 \sqrt{\frac{2}{3}}~ \varphi^{(\nu)}_{1}$&$\psi^{-(\nu)}_{-~+}$&$~=~0$~ \\[1.9 mm] 
	\rule{0pt}{6.5mm}&~$+1/2$&~$~~~0$&$\psi^{+(\nu)}_{+~0}$&$~=~-N^{(\nu)}_0 \sqrt{\frac{1}{3}}~  \varphi^{(\nu)}_{1}$&$\psi^{-(\nu)}_{+~0}$&$~=~N^{(\nu)}_0 \sqrt{\frac{1}{3}} \bigg( \frac{p^1-ip^2}{xM} \bigg)~  \varphi^{(\nu)}_{2}$~    \\[1.9 mm] 
		\rule{0pt}{6.5mm}&~$-1/2$&~$~~~0$&$\psi^{+(\nu)}_{-~0}$&$~=N^{(\nu)}_0 \sqrt{\frac{1}{3}} \bigg(\frac{p^1+ip^2}{xM} \bigg)~ \varphi^{(\nu)}_{2}$&$\psi^{-(\nu)}_{-~0}$&$~=~N^{(\nu)}_0\sqrt{\frac{1}{3}}~  \varphi^{(\nu)}_{1}$~   \\[1.9 mm]
		\rule{0pt}{6.5mm}&~$+1/2$&$~~-1$&$\psi^{+(\nu)}_{+~-}$&$~=0$&$\psi^{-(\nu)}_{+~-}$&$~=~- N^{(\nu)}_1 \sqrt{\frac{2}{3}}~  \varphi^{(\nu)}_{1}$~    \\[1.9 mm]
		\rule{0pt}{6.5mm}&~$-1/2$&$~~-1$&$\psi^{+(\nu)}_{-~-}$&$~=0$&$\psi^{-(\nu)}_{-~-}$&~$~=~N^{(\nu)}_1 \sqrt{\frac{2}{3}} \bigg(\frac{p^1+ip^2}{xM}\bigg)~  \varphi^{(\nu)}_{2}$~    \\[2.9 mm] 
		\hline
	\end{tabular}
\end{center}
	\caption{
		The LFWFs for the active quark $\lambda^{q}$ and the spectator diquark $\lambda^{Sp}$ variations of their helicities for both diquark circumstances for  $J^z=\pm1/2$. The normalization constants are $N_S$, $N^{(\nu)}_0$, and $N^{(\nu)}_1$.
	}
	\label{tab_LFWF} 
\end{table}
Derived from the predictions of soft-wall AdS/QCD \cite{Brodsky:2007hb,deTeramond:2011aml}, the general form of LFWFs $\varphi^{(\nu)}_{i}=\varphi^{(\nu)}_{i}(x,\bfp)$ listed in Table \ref{tab_LFWF} follows the parameterization $a^\nu_i,~b^\nu_i$, and $\delta^\nu$ as outlined in Ref. \cite{Maji:2016yqo}. We have
\begin{eqnarray}
	\varphi_i^{(\nu)}(x,\bfp)=\frac{4\pi}{\kappa}\sqrt{\frac{\log(1/x)}{1-x}}x^{a_i^\nu}(1-x)^{b_i^\nu}\exp\Bigg[-\delta^\nu\frac{\bfp^2}{2\kappa^2}\frac{\log(1/x)}{(1-x)^2}\bigg].
	\label{LFWF_phi}
\end{eqnarray}
\begin{table}[h]
	\centering
	\begin{tabular}{|c|c|c|}
		\hline
		\rule{0pt}{4.7mm}\backslashbox{Parameter}{$\nu$~~}       & $u$                 & $d$                         \\[0.44 mm] \hline
		\rule{0pt}{4.7mm}$C_{S}^{2}$ & $1.3872$            & $0$                     \\ \hline
		\rule{0pt}{4.7mm}$C_{V}^{2}$ & $0.6128$            & $0$                    \\[0.44 mm] \hline
		\rule{0pt}{4.7mm}$C_{VV}^{2}$ & $0$            & $1$                    \\[0.44 mm] \hline
		\rule{0pt}{4.7mm}$N_{S}$     & $2.0191$            & $0$                         \\[0.44 mm] \hline
		\rule{0pt}{4.7mm}$N_0^{\nu}$ & $3.2050$            & $5.9423$                    \\[0.44 mm] \hline
		\rule{0pt}{4.7mm}$N_1^{\nu}$ & $0.9895$            & $1.1616$                    \\[0.44 mm] \hline
		\rule{0pt}{4.7mm}$a_1^{\nu}$ & $0.280\pm 0.001$    & $0.5850 \pm 0.0003$         \\[0.44 mm] \hline
		\rule{0pt}{4.7mm}$b_1^{\nu}$ & $0.1716 \pm 0.0051$ & $0.7000 \pm 0.0002$         \\[0.44 mm] \hline
		\rule{0pt}{4.7mm}$a_2^{\nu}$ & $0.84 \pm 0.02$     & $0.9434^{+0.0017}_{-0.0013}$\\[0.44 mm] \hline
		\rule{0pt}{4.7mm}$b_2^{\nu}$ & $0.2284 \pm 0.0035$ & $0.64^{+0.0082}_{-0.0022}$  \\[0.44 mm] \hline 
		$\delta^\nu$ & $1$ & $1$   \\[0.44 mm] \hline 
	\end{tabular}
	\caption{Values of coefficients, normalization constants $N_{i}^{2}$, and model parameters corresponding to both $u$ and $d$ quarks.}
	\label{tab_par} 
\end{table}
The WFs $\varphi_i^\nu ~(i=1,2)$ happen to be distinct over the interchange $x \rightarrow 1-x$, and such asymmetry persists at the AdS/QCD limit $a_i^\nu=b_i^\nu=0$ and $\delta^\nu=1.0$ as well \cite{Gutsche:2014yea}. 
\par
The variables $a_i^{\nu}$ and $b_i^{\nu}$, appearing in Eq. \eqref{LFWF_phi}, were effectively fitted with the use of the Dirac and Pauli FFs data \cite{Maji:2016yqo,Efremov:2009ze,Burkardt:2007rv} to the model scale $\mu_0=0.313{\ \rm GeV}$. For both quark flavors, the given value of factor $\delta^{\nu}$ is assumed as the one that has been adopted from AdS/QCD \cite{deTeramond:2011aml}. Aside from this, Ref. \cite{Maji:2016yqo} is also the source of normalization constants $N_{i}^{2}$ provided in Table {\ref{tab_LFWF}}. Table \ref{tab_par} lists the model parameter values for both active quark flavors, considering the purpose of clarity. The AdS/QCD scale parameter $\kappa$, which appears in Eq. (\ref{LFWF_phi}), has been assigned a value of $0.4~\mathrm{GeV}$ \cite{Chakrabarti:2013dda,Chakrabarti:2013gra}. 
We hold the proton mass ($M$) and the constituent quark mass ($m$) to be, respectively, $0.938~\mathrm{GeV}$ and $0.055~\mathrm{GeV}$, consistent with Ref. \cite{Chakrabarti:2019wjx}.In the context of LFQDM, any physical observable $O$ for $u$ and $d$ quarks can be represented as the sum of the contributions from the diquark parts of the isoscalar-scalar ($O^{u(S)}$), isoscalar-vector ($O^{u(V)}$), and isovector-vector ($O^{d(VV)}$), are as follows
		\begin{eqnarray} 
		O^{u}&=&  ~O^{u(S)} + ~O^{u(V)}\,,\label{gtmdu} \\
		O^{d} &=&  ~O^{d(VV)}\,\label{gtmdd}.
	\end{eqnarray}

\section{{GPD correlator and parameterization at twist-$3$} \label{seccor}}
\begin{figure*}
	\centering
	\begin{minipage}[c]{0.98\textwidth}
		\includegraphics[width=0.96\textwidth]{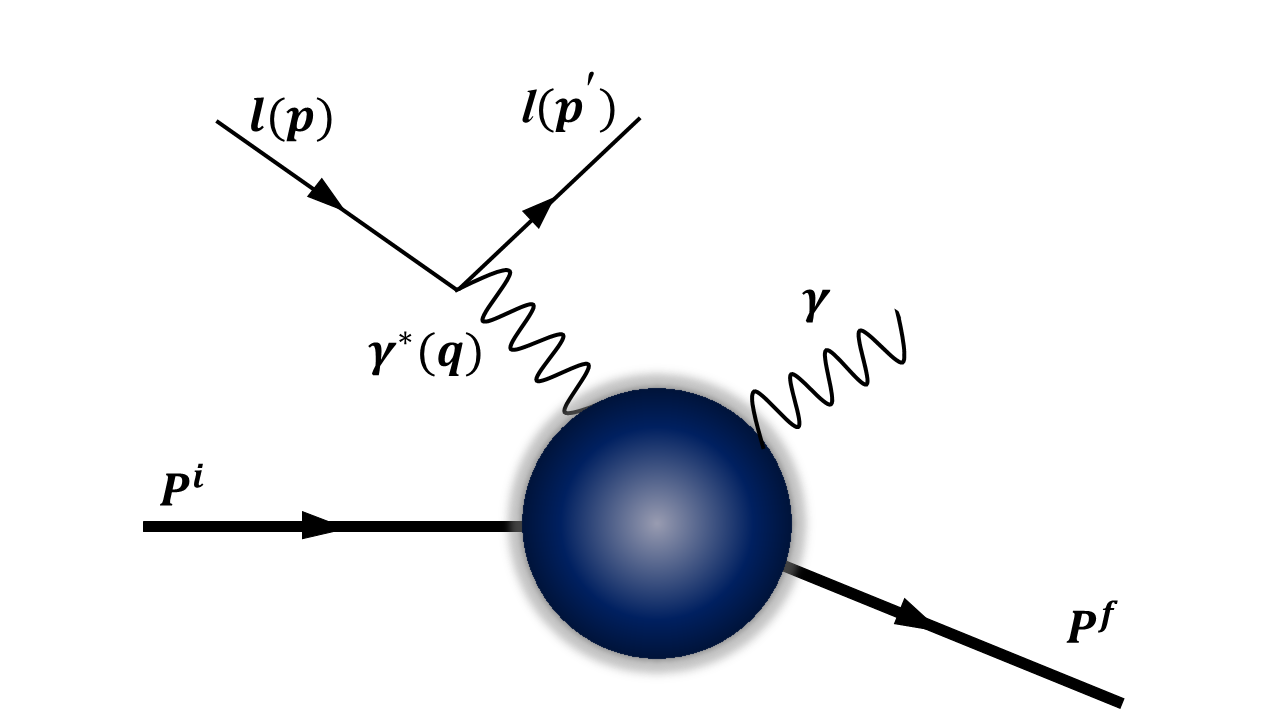}
	\end{minipage}
	\caption{\label{DVCS} Visualization of the GPDs-linked DVCS process involving a virtual photon and proton, $\gamma^* + P^i \rightarrow \gamma^* + P^f$.
	}
\end{figure*}

\noindent
This section presents a thorough examination of the GPD correlator and its parameterization. According to Ref. \cite{Meissner:2009ww}, the quark-quark GPD correlator for the proton is defined as
\begin{eqnarray} 
	F^{\nu [\Gamma]}_{[\Lambda^{N_i}\Lambda^{N_f}]}(x,\xi,t)=\frac{1}{2}\int \frac{dz^-}{2\pi} e^{\frac{i}{2}p^+ z^-} 
	\langle P^{f}; \Lambda^{N_f} |\bar{\psi} (-z/2)\Gamma \mathcal{W}_{[-z/2,z/2]} \psi (z/2) |P^{i};\Lambda^{N_i}\rangle \bigg|_{z^+=z_\perp=0}\,.
	\label{gpdcorr}
\end{eqnarray} 
In the present work, $|P^{i};\Lambda^{N_i}\rangle$ and $|P^{f};\Lambda^{N_f}\rangle$ represent the initial and final states of the proton, respectively, where $\Lambda^{N_i}$ and $\Lambda^{N_f}$ signify their helicities. The pictorial representation of the GPDs-linked DVCS process $\gamma^* + P^i \rightarrow \gamma^* + P^f$, involving a virtual photon and proton with the virtual photon is observed in the final state along with the proton, has been given in Fig. \ref{DVCS}. The GPD correlator depends on the variables set $x,\xi,$ and $t$. At zero skewness, the square of the total momentum transfer is denoted by $t= \Delta^2=-\Dp^2$, or $\xi=- \Delta^+/2P^+=0$ \cite{Meissner:2009ww}. Therefore, for the rest of the paper, we will express the GPD correlator $F^{\nu [\Gamma]}_{[\Lambda^{N_i}\Lambda^{N_f}]}(x,\xi,t)$ as $F^{\nu [\Gamma]}_{[\Lambda^{N_i}\Lambda^{N_f}]}(x,\Dp^2)$ or compactly as $F^{\nu [\Gamma]}_{[\Lambda^{N_i}\Lambda^{N_f}]}$, where $\Gamma$ stands for the twist-$3$ Dirac $\gamma$-matrices, i.e., $\Gamma=\{\gamma^j, \gamma^j\gamma_5\}$.
The Wilson line, $\mathcal{W}_{[-z/2,z/2]}$, has been considered to be $1$ for simplicity. This ensures that the related bilocal quark operator has SU$(3)$ color gauge invariance. In the present scenario, we use the convention $z^\pm=(z^0 \pm z^3)$, and we apply the symmetric frame kinematics that has been adopted in Ref. \cite{Sharma:2023ibp}. 
\par
By substituting the expression of the scalar diquark and vector diquark Fock states, Eqs.~(\ref{fockSD}) and (\ref{fockVD}), into the proton state Eq.~(\ref{PS_state}) within the GPD correlator Eq.~(\ref{gpdcorr}), one can get the GPD correlator for the scalar and vector diquark parts as overlap of LFWFs, shown in Table \ref{tab_LFWF} as
\begin{eqnarray} 
	F^{\nu [\Gamma](S)}_{[\Lambda^{N_i}\Lambda^{N_f}]}(x,\Dp^2)&=&\int\frac{C_{S}^{2}}{16\pi^3} \sum_{\lambda^{q_i}} \sum_{\lambda^{q_f}} \psi^{\Lambda^{N_f}\dagger}_{\lambda^{q_f}}\left(x,\bfp+(1-x)\frac{\Dp}{2}\right)\nonumber\\
	&&\psi^{\Lambda^{N_i}}_{\lambda^{q_i}}\left(x,\bfp-(1-x)\frac{\Dp}{2}\right) \nonumber\\  &&\frac{u^{\dagger}_{\lambda^{q_f}}\left(x P^{+},\bfp+\frac{\Dp}{2}\right)\gamma^{0} \Gamma u_{\lambda^{q_i}}\left(x P^{+},\bfp-\frac{\Dp}{2}\right)}{2 x P^{+}}{d^2 \bfp}\,, \label{cors} \\
	F^{\nu [\Gamma](A)}_{[\Lambda^{N_i}\Lambda^{N_f}]}(x,\Dp^2)&=&\int\frac{C_{A}^{2}}{16\pi^3} \sum_{\lambda^{q_i}} \sum_{\lambda^{q_f}} \sum_{\lambda^{D}} \psi^{\Lambda^{N_f}\dagger}_{\lambda^{q_f} \lambda^D}\left(x,\bfp+(1-x)\frac{\Dp}{2}\right)\nonumber\\
	&&\psi^{\Lambda^{N_i}}_{\lambda^{q_i}\lambda^D}\left(x,\bfp-(1-x)\frac{\Dp}{2}\right) \nonumber\\  &&\frac{u^{\dagger}_{\lambda^{q_f}}\left(x P^{+},\bfp+\frac{\Dp}{2}\right)\gamma^{0} \Gamma u_{\lambda^{q_i}}\left(x P^{+},\bfp-\frac{\Dp}{2}\right)}{2 x P^{+}}{d^2 \bfp}\,, \label{corv} 
\end{eqnarray} 

where $C_A = C_V, C_{VV}$ for the $u$ and $d$ quarks, respectively. The spinor product $u^{\dagger}_{\lambda^{q_f}}\left(x P^{+},\bfp+\frac{\Dp}{2}\right)$ $\gamma^{0} \Gamma u_{\lambda^{q_i}}\left(x P^{+},\bfp-\frac{\Dp}{2}\right)$ is associated with the twist-$3$ Dirac matrices. Refs. \cite{Harindranath:1996hq, Brodsky:1997de} provide a comprehensive discussion of the various Dirac spinor configurations. Here, $\lambda^{q_i}$ and $\lambda^{q_f}$ stands for initial and final states of the quark helicity, respectively. Moreover, for the vector diquark, there is an extra summation over the diquark helicity $\lambda^{D}$.
\par
Following Ref. \cite{Meissner:2009ww}, the GPDs connected to the twist-$3$ Dirac matrices $\gamma^j$ and $\gamma^j \gamma_5$ can be parameterized as
\begin{eqnarray}
 F_{[\Lambda^{N_i}\Lambda^{N_f}]}^{[\gamma^j]}
 &=& \frac{M}{2(P^+)^2} \, \bar{u}(P^{f}, \Lambda^{N_F}) \, \bigg[
      i\sigma^{+j} \, {\color{red}H_{2T}(x,\Dp^2)}
      + \frac{\gamma^+ \Dp^j  - \Delta^+ \gamma^j}{2M} \, {\color{red}E_{2T}(x,\Dp^2)} \nonumber\\*
 & & + \frac{P^+ \Dp^j  - \Delta^+ \bfP^j}{M^2} \, {\color{red}\tilde{H}_{2T}(x,\Dp^2)}
      + \frac{\gamma^+ \bfP^j  - P^+ \gamma^j}{M} \, {\color{red}\tilde{E}_{2T}(x,\Dp^2)}
     \bigg] \, u(P^{i}, \Lambda^{N_i}) \,,
 \label{parg1}\\
 F_{[\Lambda^{N_i}\Lambda^{N_f}]}^{[\gamma^j\gamma_5]}
 &=& - \frac{i\varepsilon_T^{ij} M}{2(P^+)^2} \, \bar{u}(P^{f}, \Lambda^{N_F}) \, \bigg[
      i\sigma^{+i} \, {\color{red}H'_{2T}(x,\Dp^2)}
      + \frac{\gamma^+ \Dp^i  - \Delta^+ \gamma^i}{2M} \, {\color{red}E'_{2T}(x,\Dp^2)} \nonumber\\*
 & & + \frac{P^+ \Dp^i  - \Delta^+ \bfP^i}{M^2} \, {\color{red}\tilde{H}'_{2T}(x,\Dp^2)}
      + \frac{\gamma^+ \bfP^i  - P^+ \gamma^i}{M} \, {\color{red}\tilde{E}'_{2T}(x,\Dp^2)}
     \bigg] \, u(P^{i}, \Lambda^{N_i}) . \label{parg2}
\end{eqnarray}
Here, the chiral-even GPDs are represented by functions of the form $X(x, \Dp^2)$, which are $8$ in number. The twist-$3$ chiral-even twist-$3$ GPDs have been written in \textcolor{red}{red}, while twist-$3$ GTMDs, which we refer to in the upcoming section, are written in \textcolor{blue}{blue}. In above expressions, we have used the relation $\sigma^{+ \Delta} =\sigma^{+ i}\Delta_i$ for transverse direction index i, whereas other notations have their usual meanings. 
%
\section{{Expressions of twist-$3$ chiral-even GPDs} \label{secresults}}
\noindent
To derive the expressions of the twist-$3$ chiral-even GPDs for each kind of diquark, we have substituted the proton state Eq. \eqref{PS_state} with proper polarization in the correlator Eq. \eqref{gpdcorr} via scalar and vector diquark Fock states from Eq. \eqref{fockSD} and Eq. \eqref{fockVD}, respectively.  One can obtain certain twist-$3$ chiral-even GPD by choosing matrix structure $\Gamma= \gamma^j$ from Eq. {\eqref{parg1}}
	\begin{eqnarray}
\frac{i \Dp^2}{P^+} {\color{red}\tilde{E}_{2T}^{\nu}}
	 &=&\Delta_y \left(F^{\nu[\gamma^1]}_{[++]} -F^{\nu[\gamma^1]}_{[--]}\right)-\Delta_x \left(F^{\nu[\gamma^2]}_{[++]} -F^{\nu[\gamma^2]}_{[--]} \right), \label{g1.1}\\
 \frac{-2 i M \Dp^2}{ P^+} {\color{red}H_{2T}^{\nu}}	&=& \left( \Delta_x + i \Delta_y\right)	
	\left(\Delta_y~F^{\nu[\gamma^1]}_{[-+]} -\Delta_x F^{\nu[\gamma^2]}_{[-+]} \right) 
 \nonumber\\
		&& \qquad + \left( \Delta_x - i \Delta_y\right)	\left( \Delta_y~F^{\nu[\gamma^1]}_{[+-]} -\Delta_x F^{\nu[\gamma^2]}_{[+-]} \right) ,
 \label{g1.2}\\
 \frac{4 M^2 {\color{red}H_{2T}^{\nu}}+ {\color{red}\tilde{H}_{2T}^{\nu}}\Dp^2}{M P^+} &=&	 \left( ~F^{\nu[\gamma^1]}_{[-+]} -F^{\nu[\gamma^1]}_{[+-]} \right) 
	+ i 
	\left( ~F^{\nu[\gamma^2]}_{[-+]} +F^{\nu[\gamma^2]}_{[+-]} \right),
 \label{g1.3}\\
 \frac{2 \left(\bfp \cdot \Dp \right){\color{red}\tilde{E}_{2T}^{\nu}}+\left({\color{red}E_{2T}^{\nu}}+2{\color{red}\tilde{H}_{2T}^{\nu}}\right)\Dp^2}{ P^+}&=&	\Delta_x \left(~F^{\nu[\gamma^1]}_{[++]} +F^{\nu[\gamma^1]}_{[--]}\right)+\Delta_y \left( F^{\nu[\gamma^2]}_{[++]} +F^{\nu[\gamma^2]}_{[--]} \right) \label{g1.4}.
  \end{eqnarray}
  Similarly, for the matrix structure  $\Gamma=\gamma^j\gamma_5$, we have obtained the following equations
  \begin{eqnarray}
   \frac{\Dp^2}{ P^+} {\color{red}\tilde{E}_{2T}^{'\nu}}
	&=&\Delta_x \left(F^{\nu[\gamma^1\gamma_5]}_{[++]} -F^{\nu[\gamma^1 \gamma_5]}_{[--]}\right)+\Delta_y  \left(F^{\nu[\gamma^2 \gamma_5]}_{[++]} -F^{\nu[\gamma^2 \gamma_5]}_{[--]}\right) , \label{r1.5}\\
 \frac{2 M \Dp^2}{P^+}{\color{red}H_{2T}^{'\nu}}  &=&  \left( \Delta_x + i \Delta_y\right)	
	\left( \Delta_x~W^{\nu[\gamma^1 \gamma_5]}_{[-+]} +\Delta_y F^{\nu[\gamma^2 \gamma_5]}_{[-+]} \right) 
	 \nonumber\\
		&& \qquad + \left( \Delta_x - i \Delta_y\right)	\left( \Delta_x~F^{\nu[\gamma^1 \gamma_5]}_{[+-]} \Delta_y F^{\nu[\gamma^2 \gamma_5]}_{[+-]} \right),
 \label{g1.6}\\
  \frac{4 M^2 {\color{red}H_{2T}^{'\nu}}+ {\color{red}\tilde{H}_{2T}^{'\nu}}\Dp^2}{M P^+}	&=& \left(F^{\nu[\gamma^1 \gamma_5]}_{[-+]} -F^{\nu[\gamma^1\gamma_5]}_{[+-]} \right) + i \left(F^{\nu[\gamma^2\gamma_5]}_{[-+]} -F^{\nu[\gamma^2\gamma_5]}_{[+-]} \right)  ,\label{g1.7}\\
  i\frac{2 \left(\bfp \cdot \Dp \right){\color{red}\tilde{E}_{2T}^{'\nu}}+\left({\color{red}E_{2T}^{'\nu}}+2{\color{red}\tilde{H}_{2T}^{'\nu}}\right)\Dp^2}{ P^+} &=&
 \Delta_y \left(F^{\nu[\gamma^1 \gamma_5]}_{[++]} +F^{\nu[\gamma^1 \gamma_5]}_{[--]}\right)-\Delta_x \left( F^{\nu[\gamma^2 \gamma_5]}_{[++]} +F^{\nu[\gamma^2 \gamma_5]}_{[--]} \right). \label{g1.8}
 \end{eqnarray}
	We define
	\begin{eqnarray}
		T_{ij}^{(\nu)}(x,\bfp,\Dp)&=&\varphi_i^{(\nu) \dagger}\left(x,\bfp+(1-x)\frac{\Dp}{2}\right) \varphi_j^{(\nu)}\left(x,\bfp-(1-x)\frac{\Dp}{2}\right)
		\label{Tij1},
	\end{eqnarray}
where, $i,j=1,2$. By using the wave function from Eq. \eqref{LFWF_phi} with the aforementioned equation, one may infer
	\begin{eqnarray}
		T_{ij}^{(\nu)}(x,\bfp,\Dp)&=&T_{ji}^{(\nu)}(x,\bfp,\Dp)\label{Tij2},\\
		\varphi_i^{(\nu)\dagger}\left(x,\bfp+(1-x)\frac{\Dp}{2}\right)&=&\varphi_i^{(\nu)}\left(x,\bfp+(1-x)\frac{\Dp}{2}\right)\label{Tij3}.
	\end{eqnarray}
For the twist-$3$ Dirac matrix structure, the chiral-even GPD expressions for both diquark possibilities can be written as
	\begin{eqnarray} 
		{\color{red}x \tilde{E}_{2T}^{\nu(S)}} &=& \int  \frac{C_{S}^{2} N_s^2}{16 \pi^3} \left(- T_{11}^{\nu}+ \left( \left(\bfp^2-(1-x)^2\frac{\Dp^2}{4}\right)-2\left(\frac{\bfp^2 \Dp^2 - (\bfp \cdot \Dp)^2}{\Dp^2} \right)(1-x ) \right)\frac{T_{22}^{\nu}}{x^2 M^2}\right) ~{d^2 \bfp}, \nonumber\\
\label{xE2TBs}\\
		%
		{\color{red}x \tilde{E}_{2T}^{\nu(A)}} &=& \int  \frac{C_{A}^{2}}{16 \pi^3}  \bigg(\frac{1}{3} |N_0^\nu|^2-\frac{2}{3}|N_1^\nu|^2 \bigg) \Bigg(- T_{11}^{\nu}+ \Bigg( \left(\bfp^2-(1-x)^2\frac{\Dp^2}{4}\right) \nonumber\\	&&-2\left(\frac{\bfp^2 \Dp^2 - (\bfp \cdot \Dp)^2}{\Dp^2} \right)(1-x ) \Bigg)\frac{T_{22}^{\nu}}{x^2 M^2}\Bigg) ~{d^2 \bfp}, \label{xE2TBv}
	\end{eqnarray}
		\begin{eqnarray} 
		%
			{\color{red}xH_{2T}^{'\nu(S)}} &=&\int  \frac{C_{S}^{2} N_s^2}{16 \pi^3} \frac{1}{M} \bigg(m T_{11}^{\nu}+ 2\frac{(\bfp \cdot \Dp)^2}{\Dp^2} \frac{T_{12}^{\nu}}{xM}-m  \Bigg( \Bigg(\frac{2(\bfp \cdot \Dp)^2-\bfp^2 \Dp^2}{\Dp^2} \Bigg)\nonumber\\
		&&-(1-x)^2\frac{\Dp^2}{4} \Bigg)\frac{T_{22}^{\nu}}{x^2 M^2}\bigg) ~{d^2 \bfp} , \label{xH2TPs}\\
		%
		%
		{\color{red}xH_{2T}^{'\nu(A)}} &=& \int -\frac{C_{A}^{2} }{16 \pi^3} \bigg(\frac{1}{3} |N_0^\nu|^2\bigg) \frac{1}{M} \bigg(m T_{11}^{\nu}+ 2\frac{(\bfp \cdot \Dp)^2}{\Dp^2} \frac{T_{12}^{\nu}}{xM}-m  \Bigg( \Bigg(\frac{2(\bfp \cdot \Dp)^2-\bfp^2 \Dp^2}{\Dp^2} \Bigg)\nonumber\\
		&&-(1-x)^2\frac{\Dp^2}{4} \Bigg)\frac{T_{22}^{\nu}}{x^2 M^2}\bigg) ~{d^2 \bfp} , \label{xH2TPv}\\
		%
		{\color{red}x \tilde{H}_{2T}^{'\nu(S)}} &=& \int \frac{C_{S}^{2} N_s^2}{4\pi^3}\frac{M}{\Dp^2}  \Bigg[\Bigg(\Bigg(\frac{\bfp^2 \Dp^2-2(\bfp \cdot \Dp)^2}{\Dp^2} \Bigg)+(1-x)\frac{\Dp^2}{4}\Bigg) \frac{T_{12}^{\nu}}{xM}-m\Bigg(\Bigg(\frac{\bfp^2 \Dp^2-2(\bfp \cdot \Dp)^2}{\Dp^2} \Bigg)\nonumber\\
		&&+(1-x)^2\frac{\Dp^2}{4}\Bigg)\frac{T_{22}^{\nu}}{x^2M^2}\Bigg]~{d^2 \bfp}, \label{xH2TBPs}
		\\
		%
		%
		{\color{red}x \tilde{H}_{2T}^{'\nu(A)}} &=& \int -\frac{C_{A}^{2}}{4 \pi^3} \bigg(\frac{1}{3} |N_0^\nu|^2\bigg)\frac{M}{\Dp^2}  \Bigg[\Bigg(\Bigg(\frac{\bfp^2 \Dp^2-2(\bfp \cdot \Dp)^2}{\Dp^2} \Bigg)+(1-x)\frac{\Dp^2}{4}\Bigg) \frac{T_{12}^{\nu}}{xM}-m\Bigg(\Bigg(\frac{\bfp^2 \Dp^2-2(\bfp \cdot \Dp)^2}{\Dp^2} \Bigg)\nonumber\\
		&&+(1-x)^2\frac{\Dp^2}{4}\Bigg)\frac{T_{22}^{\nu}}{x^2M^2}\Bigg]~{d^2 \bfp} ,  \label{xH2TBPv}\\
		%
		%
		%
		{\color{red}xE_{2T}^{'\nu(S)}} &=&  \int \frac{C_{S}^{2} N_s^2}{4\pi^3}  \bigg(-T_{11}^{\nu}+2m\left(1-x\right) \frac{T_{12}^{\nu}}{x M}+2 \bigg( \frac{\bfp^2 \Dp^2 - (\bfp \cdot \Dp)^2}{\Dp^2}  \bigg)\frac{T_{22}^{\nu}}{x^2 M^2} - \bigg( \bfp^2-(1-x)^2\frac{\Dp^2}{4} \bigg)\frac{T_{22}^{\nu}}{x^2 M^2}\bigg)\nonumber\\
		&& -2 \int \frac{C_{S}^{2} N_s^2}{4\pi^3}\frac{M}{\Dp^2}  \Bigg[\Bigg(\Bigg(\frac{\bfp^2 \Dp^2-2(\bfp \cdot \Dp)^2}{\Dp^2} \Bigg)+(1-x)\frac{\Dp^2}{4}\Bigg) \frac{T_{12}^{\nu}}{xM}-m\Bigg(\Bigg(\frac{\bfp^2 \Dp^2-2(\bfp \cdot \Dp)^2}{\Dp^2} \Bigg)\nonumber\\
		&&+(1-x)^2\frac{\Dp^2}{4}\Bigg)\frac{T_{22}^{\nu}}{x^2M^2}\Bigg]~{d^2 \bfp} ,  \label{xE2TPs}\\
		%
		%
		{\color{red}xE_{2T}^{'\nu(A)}} &=&  \int \frac{C_{A}^{2} }{4\pi^3}\bigg(\frac{1}{3} |N_0^\nu|^2+\frac{2}{3}|N_1^\nu|^2 \bigg) 
	 \bigg(-T_{11}^{\nu}+2m\left(1-x\right) \frac{T_{12}^{\nu}}{x M}+ 2\bigg( \frac{\bfp^2 \Dp^2 - (\bfp \cdot \Dp)^2}{\Dp^2}  \bigg)\frac{T_{22}^{\nu}}{x^2 M^2} - \nonumber\\
	 &&\bigg( \bfp^2-(1-x)^2\frac{\Dp^2}{4} \bigg)\frac{T_{22}^{\nu}}{x^2 M^2}\bigg) +2\int \frac{C_{A}^{2}}{4 \pi^3} \bigg(\frac{1}{3} |N_0^\nu|^2\bigg)\frac{M}{\Dp^2}  \Bigg[\Bigg(\Bigg(\frac{\bfp^2 \Dp^2-2(\bfp \cdot \Dp)^2}{\Dp^2} \Bigg)+(1-x)\frac{\Dp^2}{4}\Bigg) \frac{T_{12}^{\nu}}{xM}\nonumber\\
	 &&-m\Bigg(\Bigg(\frac{\bfp^2 \Dp^2-2(\bfp \cdot \Dp)^2}{\Dp^2} \Bigg)+(1-x)^2\frac{\Dp^2}{4}\Bigg)\frac{T_{22}^{\nu}}{x^2M^2}\Bigg]~{d^2 \bfp} . \label{xE2TPv}
	\end{eqnarray}
It should be noted that GPDs $xE_{2T}^{\nu}$, $x H_{2T}^{\nu}$, and $x\tilde{H}_{2T}^{\nu}$ concerining with matrix strucrure $ \gamma^j$  and  $x\tilde{E}_{2T}^{'\nu}$ concerning the matrix structure $\gamma^j \gamma_5$ were found to vanish in our calculation. This result is in line with the BLFQ findings \cite{Zhang:2023xfe}. 
Also, the Lattice QCD computation of twist-$3$ chiral-even axial-vector GPD  $x \tilde{E}_{2T}^{'\nu}$ comes out to be $0$, which is in sync with our calculations \cite{Bhattacharya:2023jsc}.
%
\section{Relation with twist-$3$ GTMDs} \label{secrel1}
\noindent
Understanding the GTMD correlator structure and its parameterization equations is necessary to ascertain the relationship between GPDs and GTMDs. The GPD correlator for zero skewness $F^{\nu [\Gamma]}_{[\Lambda^{N_i}\Lambda^{N_f}]}(x,\Delta_\perp^2)$ can be expressed in terms of the fully unintegrated quark-quark GTMD correlator $W^{\nu [\Gamma]}_{[\Lambda^{N_i}\Lambda^{N_f}]}(x, \mathbf{p}_\perp^2, \Delta_\perp^2, \mathbf{p}_\perp \cdot \Delta_\perp)$ as  \cite{Meissner:2009ww}
\begin{eqnarray} 
F^{\nu [\Gamma]}_{[\Lambda^{N_i}\Lambda^{N_f}]}(x,\Dp^2)=\int {d^2 \bfp} 
	W^{\nu [\Gamma]}_{[\Lambda^{N_i}\Lambda^{N_f}]}(x, \bfp^2,\Dp^2,\bfp \cdot \Dp).
	\label{gtmdcorr}
\end{eqnarray} 
According to Ref. \cite{Meissner:2009ww}, the quark GTMDs can be presented for different Dirac matrix structure values $\Gamma= \gamma^j$ and $\gamma^j\gamma_5$ as
\begin{eqnarray} %
W_{[\Lambda^{N_i}\Lambda^{N_f}]}^{[\gamma^{j}]}
		&=& \frac{1}{2P^+} \, \bar{u}(P^{f}, \Lambda^{N_F}) \, \bigg[
		\frac{ p_{\perp}^j}{M} \,	{\color{blue} F_{2,1}}
		+ \frac{\Delta_{\perp}^j}{M} \,	{\color{blue} F_{2,2}}
		+ \frac{M \, i\sigma^{j+}}{k^+} \,	{\color{blue} F_{2,3}} \nonumber\\*
		& &  + \frac{p_{\perp}^j \, i\sigma^{~\rho +} p_{\perp}^{~\rho}}{M \, P^+} \,	{\color{blue}F_{2,4}}
		+ \frac{{\Delta_\perp}^j \, i\sigma^{~\rho +} p_T^{~\rho}}{M \, P^+} \,	{\color{blue} F_{2,5}}
		+ \frac{\Delta_{\perp}^j \, i\sigma^{~\rho +} \Delta_{\perp}^\rho}{M \, P^+} \,	{\color{blue} F_{2,6}} \nonumber\\*
		& &  + \frac{p_{\perp}^i \, i\sigma^{ij}}{M} \, 	{\color{blue}F_{2,7}}
		+ \frac{\Delta_{\perp}^i \, i\sigma^{ij}}{M} \, 	{\color{blue}F_{2,8}}
		\bigg] \, u(P^{i}, \Lambda^{N_i})
		\,,   	\\ 	W_{[\Lambda^{N_i}\Lambda^{N_f}]}^{[\gamma^{j}\gamma_5]}
		&=& \frac{1}{2P^+} \, \bar{u}(P^{f}, \Lambda^{N_F}) \, \bigg[
		- \frac{i\varepsilon_{\perp}^{ij} p_{\perp}^i}{M} \, 	{\color{blue}G_{2,1}}
		- \frac{i\varepsilon_{\perp}^{ij} \Delta_{\perp}^i}{M} \, 	{\color{blue}G_{2,2}}
		+ \frac{M \, i\sigma^{j+}\gamma_5}{P^+} \, 	{\color{blue}G_{2,3}}\nonumber
	\\*
 	& &  + \frac{p_{\perp}^j \, i\sigma^{~\rho +}\gamma_5 p_{\perp}^{~\rho}}{M \, P^+} \,	{\color{blue} G_{2,4}}
		+ \frac{{\Delta_\perp}^j \, i\sigma^{~\rho +}\gamma_5 p_{\perp}^{~\rho}}{M \, P^+} \,	{\color{blue} G_{2,5}}
		+ \frac{{\Delta_\perp}^j \, i\sigma^{~\rho +}\gamma_5 \Delta_{\perp}^{~\rho}}{M \, P^+} \, 	{\color{blue}G_{2,6}} \nonumber\\*
		& &  + \frac{p_{\perp}^j \, i\sigma^{+-}\gamma_5}{M} \, 	{\color{blue}G_{2,7}}
		+ \frac{\Delta_{\perp}^j \, i\sigma^{+-}\gamma_5}{M} \, 	{\color{blue}G_{2,8}}
		\bigg] \, u(P^{i}, \Lambda^{N_i})\,.
\end{eqnarray}
On solving the twist-$3$ GTMD parametrization parallel to the 
Eqs. (\ref {g1.1}) to (\ref {g1.8}), we get the following relations for Dirac matrix structure $\gamma^j$ 
\begin{eqnarray}
 \frac{i P^+}{\Dp^2}\Bigg(\Delta_y \left( ~W^{\nu[\gamma^1]}_{[++]} -W^{\nu[\gamma^1]}_{[--]}\right)-\Delta_x \left(W^{\nu[\gamma^2]}_{[++]} -W^{\nu[\gamma^2]}_{[--]}\right) \Bigg)&=& 2~\bigg[
	\frac{\bfp \cdot \Dp}{\Dp^2} \, {\color{blue}F_{2,7}} 
	+ {\color{blue}F_{2,8}} 
	\bigg]
	, \label{r1.1}\\
 \frac{i P^+}{2 M \Dp^2} 
	\Bigg( \left( \Delta_x + i \Delta_y\right)	
	\left( \Delta_y~W^{\nu[\gamma^1]}_{[-+]} -\Delta_x W^{\nu[\gamma^2]}_{[-+]} \right) 
 &=&\bigg[-{\color{blue}F_{2,3}} \nonumber \\ 
 + \left( \Delta_x - i \Delta_y\right)	
	\left( \Delta_y~W^{\nu[\gamma^1]}_{[+-]} -\Delta_x W^{\nu[\gamma^2]}_{[+-]} \right)
	\Bigg) &&+ \frac{(\bfp \cdot \Dp)^2-\bfp^2 \Dp^2}{M^2 \Dp^2} {\color{blue}F_{2,4}}  \bigg] ,
 \label{r1.2}\\
	 M P^+\Bigg(
	\left( ~W^{\nu[\gamma^1]}_{[-+]} -W^{\nu[\gamma^1]}_{[+-]} \right) 
	+ i 
	\left( ~W^{\nu[\gamma^2]}_{[-+]} +W^{\nu[\gamma^2]}_{[+-]} \right)
	\Bigg) &=& \bigg[(\bfp \cdot \Dp){\color{blue}F_{2,1}} + 
\Dp^2 {\color{blue}F_{2,2}} \nonumber \\ 
 &&-4 M^2 {\color{blue}F_{2,3}} - \bfp^2 {\color{blue}F_{2,4}} \nonumber \\ 
 &&-2 (\bfp \cdot \Dp) {\color{blue}F_{2,5}} - 2 \Dp^2 {\color{blue}F_{2,6}} \bigg] ,
 \label{r1.3}\\
P^+\Bigg(\Delta_x \left( ~W^{\nu[\gamma^1]}_{[++]} +W^{\nu[\gamma^1]}_{[--]}\right)+\Delta_y \left( W^{\nu[\gamma^2]}_{[++]} +W^{\nu[\gamma^2]}_{[--]} \right)\Bigg) &=&2\bigg[(\bfp \cdot \Dp){\color{blue}F_{2,1}} + 
\Dp^2 {\color{blue}F_{2,2}}	\bigg].\label{r1.4}
 \end{eqnarray}
Similarly, for the Dirac matrix structure  $\Gamma=\gamma^j\gamma_5$
\begin{eqnarray}
 \frac{ P^+}{\Delta^2_{\perp}}\Bigg(\Delta_x \left( ~W^{\nu[\gamma^1\gamma_5]}_{[++]} -W^{\nu[\gamma^1 \gamma_5]}_{[--]}\right)+\Delta_y  \left( W^{\nu[\gamma^2 \gamma_5]}_{[++]} -W^{\nu[\gamma^2 \gamma_5]}_{[--]}\right) \Bigg)&=& 4~\bigg[
	\frac{\bfp \cdot \Dp}{\Dp^2} \, {\color{blue}G_{2,7}} 
	+ {\color{blue}G_{2,8}} 
	\bigg]
	, \label{r1.5}\\
 \frac{ P^+}{2 M \Dp^2} 
	\Bigg( \left( \Delta_x + i \Delta_y\right)	
	\left( \Delta_x~W^{\nu[\gamma^1 \gamma_5]}_{[-+]} +\Delta_y W^{\nu[\gamma^2 \gamma_5]}_{[-+]} \right) 
 &=&\bigg[{\color{blue}G_{2,3}} +\frac{\Dp^2}{M^2}\bigg(\frac{(\bfp \cdot \Dp)^2}{\Dp^4} {\color{blue}G_{2,4}} \nonumber \\ 
 + \left( \Delta_x - i \Delta_y\right)	
	\left( \Delta_x~W^{\nu[\gamma^1 \gamma_5]}_{[+-]} \Delta_y W^{\nu[\gamma^2 \gamma_5]}_{[+-]} \right)
	\Bigg) &&+\frac{(\bfp \cdot \Dp)}{\Dp^2} {\color{blue}G_{2,5}} +{\color{blue}G_{2,6}}\bigg) \bigg] ,
 \label{r1.6}\\
 	 M P^+\Bigg(	\left(W^{\nu[\gamma^1 \gamma_5]}_{[-+]} -W^{\nu[\gamma^1\gamma_5]}_{[+-]} \right) + i \left(W^{\nu[\gamma^2\gamma_5]}_{[-+]} -W^{\nu[\gamma^2\gamma_5]}_{[+-]} \right)\Bigg) &=& \bigg[(\bfp \cdot \Dp){\color{blue}G_{2,1}} + \Dp^2 {\color{blue}G_{2,2}}\nonumber \\ 
 &&-4 M^2 {\color{blue}G_{2,3}} +2\bfp^2 {\color{blue}G_{2,4}} \nonumber \\ 
 &&+2 (\bfp \cdot \Dp) {\color{blue}G_{2,5}} + 2 \Dp^2 {\color{blue}G_{2,6}} \bigg] ,\label{r1.7}\\
 P^+\Bigg(\Delta_y \left( ~W^{\nu[\gamma^1 \gamma_5]}_{[++]} +W^{\nu[\gamma^1 \gamma_5]}_{[--]}\right)-\Delta_x \left( W^{\nu[\gamma^2 \gamma_5]}_{[++]} +W^{\nu[\gamma^2 \gamma_5]}_{[--]} \right)\Bigg) &=& 2 i\bigg[(\bfp \cdot \Dp){\color{blue}G_{2,1}} + \Dp^2 {\color{blue}G_{2,2}}\bigg].\label{r1.8}
 \end{eqnarray}
 To obtain the GPD relations with GTMDs, we have
compared Eqs. (\ref{g1.1}) to (\ref{g1.8})  with Eqs. (\ref{r1.1}) to (\ref{r1.8})  via the use of Eq. (\ref{gtmdcorr})

\begin{eqnarray}
\color{red}{H_{2T}(x,\xi,t)} &=& \int {d^2 \bfp} \, \bigg[
-{\color{blue}F_{2,3}}
+ \frac{(\bfp \cdot \Dp)^2 - \bfp^2 \Dp^2}{M^2  \Dp^2} \, {\color{blue}F_{2,4}}
\bigg] , \label{e:gpd_gtmd_17} \\
{\color{red}E_{2T}(x,\xi,t)} &=& \int {d^2 \bfp} \, \bigg[
4 \bigg(
\frac{2 (\bfp \cdot \Dp)^2 - \bfp^2 \Dp^2}{(\Dp^2)^2} \, {\color{blue}F_{2,4}}
+ \frac{(\bfp \cdot \Dp)}{\Dp^2} \, {\color{blue}F_{2,5}}+ {\color{blue}F_{2,6}}
\bigg)\nonumber\\* &&~-~4\bigg(
\frac{(\bfp \cdot \Dp)^2}{\Dp^2} \, {\color{blue}F_{2,7}}
+ (\bfp \cdot \Dp) {\color{blue}F_{2,8}}
\bigg)
\bigg] \, \label{e:gpd_gtmd_18} \\
{\color{red}\tilde{H}_{2T}(x,\xi,t)} &=& \int {d^2 \bfp} \, \bigg[\bigg(\frac{(\bfp \cdot \Dp)}{\Dp^2} \, {\color{blue}F_{2,1}}
+ {\color{blue}F_{2,2}}
\bigg) \nonumber\\* && \qquad\qquad
- 2 \bigg(
\frac{2 (\bfp \cdot \Dp)^2 - \bfp ^2 \Dp^2}{(\Dp^2)^2} \, {\color{blue}F_{2,4}}
+ \frac{\bfp \cdot \Dp}{\Dp^2} \, {\color{blue}F_{2,5}}
+ {\color{blue}F_{2,6}}
\bigg) \bigg] \,, \label{e:gpd_gtmd_19} \\
{\color{red}\tilde{E}_{2T}(x,\xi,t)} &=& \int {d^2 \bfp}\, \bigg[
- 2 \bigg(
\frac{(\bfp \cdot \Dp)}{\Dp^2} \, {\color{blue}F_{2,7}}
+ {\color{blue}F_{2,8}}
\bigg)
\bigg] , \label{e:gpd_gtmd_20}\\
{\color{red}H'_{2T}(x,\xi,t)} &=& \int {d^2 \bfp} \, \bigg[
{\color{blue}G_{2,3}}
+ \frac{\Dp^2}{M^2} \bigg(
\frac{(\bfp \cdot \Dp)^2}{(\Dp^2)^2} \, {\color{blue}G_{2,4}}
+ \frac{\bfp \cdot \Dp}{\Dp^2} \, {\color{blue}G_{2,5}}
+ {\color{blue}G_{2,6}}
\bigg)
\bigg] \, \label{e:gpd_gtmd_21} \\
{\color{red}E'_{2T}(x,\xi,t)} &=& \int {d^2 \bfp} \, \bigg[
4 \bigg(
\frac{2 (\bfp \cdot \Dp)^2 - \bfp^2 \Dp^2}{(\Dp^2)^2} \, {\color{blue}G_{2,4}}
+ \frac{\bfp \cdot \Dp}{\Dp^2} \, {\color{blue}G_{2,5}}
+ {\color{blue}G_{2,6}}
\bigg)
\bigg] \, \label{e:gpd_gtmd_22} \\
{\color{red}\tilde{H}'_{2T}(x,\xi,t)} &=& \int {d^2 \bfp} \, \bigg[
\bigg(
\frac{\bfp \cdot \Dp}{\Dp^2} \, {\color{blue}G_{2,1}}
+ {\color{blue}G_{2,2}}
\bigg) \nonumber\\* && \qquad\qquad
- 2 \bigg(
\frac{2 (\bfp \cdot \Dp)^2 - \bfp ^2 \Dp^2}{(\Dp^2)^2} \, {\color{blue}G_{2,4}}
+ \frac{\bfp \cdot \Dp}{\Dp^2} \, {\color{blue}G_{2,5}}
+ {\color{blue}G_{2,6}}
\bigg) 
\bigg] \,, \label{e:gpd_gtmd_23} \\
{\color{red}\tilde{E}'_{2T}(x,\xi,t)} &=& \int {d^2 \bfp} \, \bigg[ 4 \bigg(
\frac{\bfp \cdot \Dp}{\Dp^2} \, {\color{blue}G_{2,7}}
+ {\color{blue}G_{2,8}}
\bigg)
\bigg] \, . \label{e:gpd_gtmd_24}\end{eqnarray}
The twist-$3$ GPDs-GTMDs relations mentioned above have been satisfied by our calculations, similar to these have been suggested in Ref. \cite{Meissner:2009ww}. Contrary to that, in our exploration, we find that the relation given in Eq. \eqref{e:gpd_gtmd_18} and Eq. \eqref{e:gpd_gtmd_24} is not followed. In Eq. \eqref{e:gpd_gtmd_18} relation, the GTMDs $F_{2,7}$ and $F_{2,8}$ appear whose contribution was found to be negligible while for the relation given in Eq. \eqref{e:gpd_gtmd_24} a factor of $2$ gets multiplied to the right-hand side.
\section{Relation with twist-$3$ TMDs} \label{secrel2}
\noindent
By the application of zero momentum tansfer limit on GTMDs $G_{2,3}$ and $G_{2,4}$ , we can get the twist-$3$ TMDs $g'_T(x,\bfp^2)$ and $g_{T}^\bot(x,\bfp^2)$ sequentially. So, for $\Dp=0$ Eq. \eqref{e:gpd_gtmd_22} is reduced to the following relation 
\begin{eqnarray}
    H'_{2T}(x,0,0)
 &=& \int {d^2 \bfp} \, \bigg[ G_{2,3}^e(x,0,\bfp^2,0,0)
     + \frac{\bfp^2}{2 M^2} \, G_{2,4}^e(x,0,\bfp^2,0,0) \bigg] \nonumber\\*
 &=& \int {d^2 \bfp} \, \bigg[ g'_T(x,\bfp^2)
     + \frac{\bfp^2}{2 M^2} \, g_{T}^\bot(x,\bfp^2) \bigg]\,. 
\end{eqnarray}
This relation can also written as $   g_{T}(x) = \lim \limits_{\Delta \rightarrow 0} H'_{2T}(x,0,-t) $, where is the $g_{T}(x)$ is PDF obtained from the concerned TMD and has also been verified for BLFQ in Ref. \cite{Zhang:2023xfe}.

\section{{Discussion}\label{secdiscussion}}
\noindent
This section presents the numerical results for twist-$3$ chiral-even GPDs of the proton in the LFQDM, focusing on the Dirac matrix structures $\gamma^j$ and $\gamma^j \gamma_5$ at zero skewness. The discussion encompasses the twist-$3$ chiral-even GPDs, the twist-$3$ IPDPDFs, and the twist-$3$ FFs in the following subsections.
\subsection{GPDs}\label{ssgpds}
\begin{figure*}
	\centering
	\begin{minipage}[c]{1\textwidth}
		(a)\includegraphics[width=0.44\textwidth]{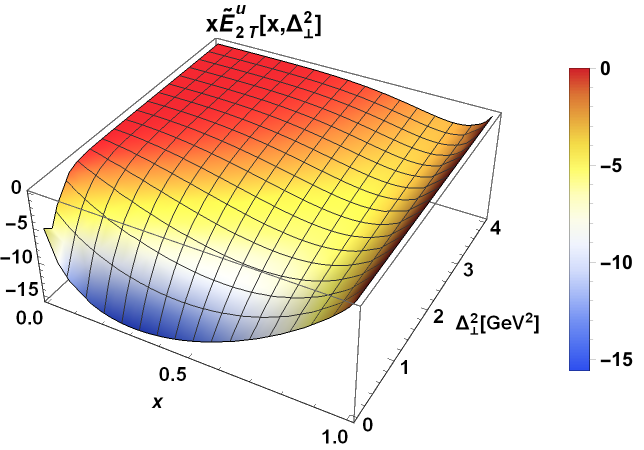}
		\hspace{0.05cm}
		(b)\includegraphics[width=0.44\textwidth]{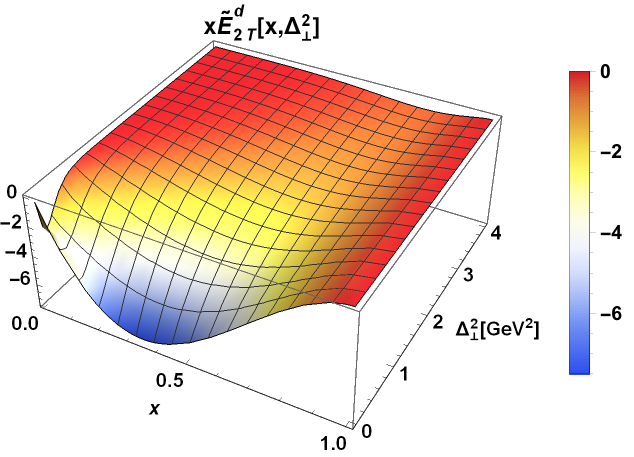}\\
		(c)\includegraphics[width=0.44\textwidth]{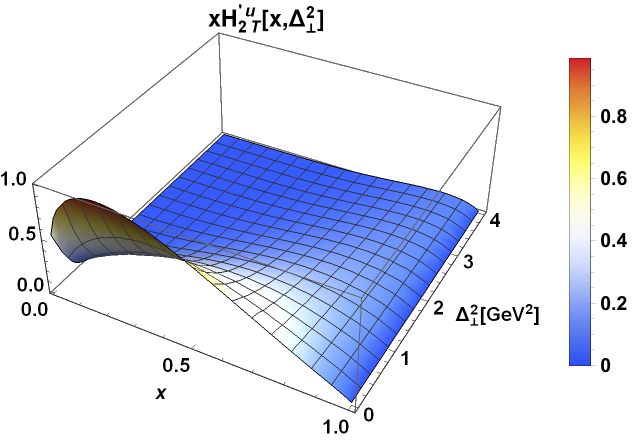}
		\hspace{0.05cm}
		(d)\includegraphics[width=0.44\textwidth]{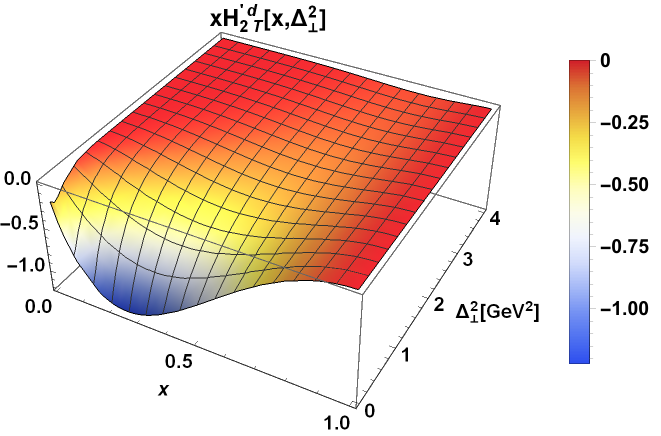}\\
		(e)\includegraphics[width=0.44\textwidth]{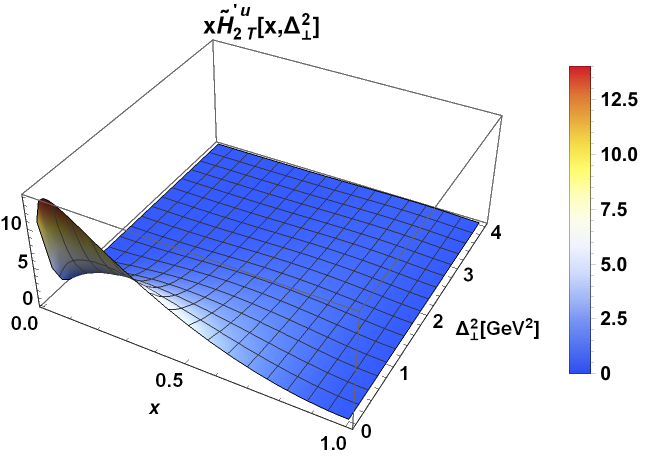}
		\hspace{0.05cm}
		(f)\includegraphics[width=0.44\textwidth]{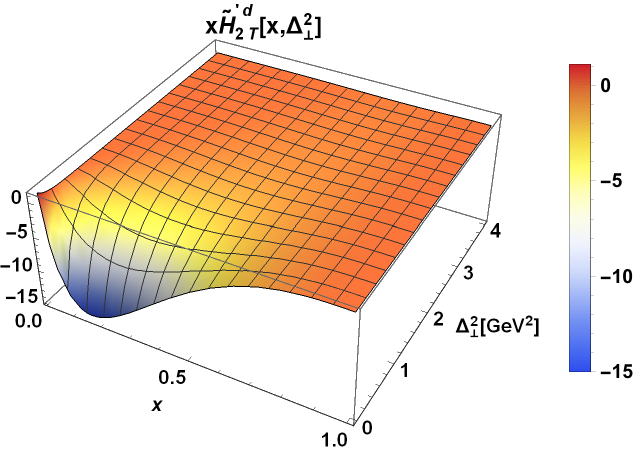}\\
		(g)\includegraphics[width=0.44\textwidth]{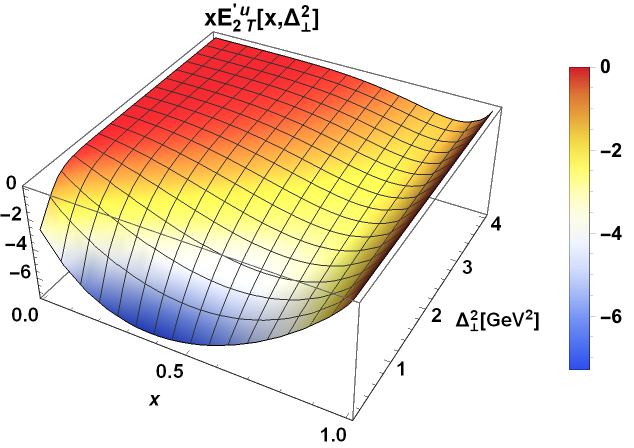}
		\hspace{0.05cm}
		(h)\includegraphics[width=0.44\textwidth]{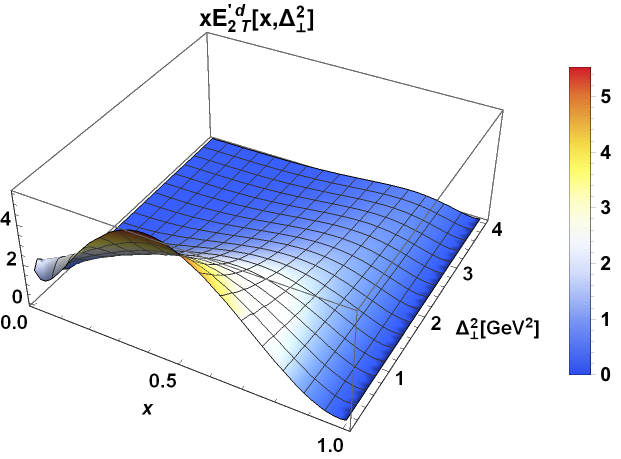}
		\\
	\end{minipage}
	\caption{\label{fig3dvxd}  Plots of the chiral-even twist-$3$ GPDs $x \tilde{E}_{2T}^{\nu} $, $ x H_{2T}^{'\nu} $, $ x \tilde{H}_{2T}^{'\nu} $, and $ x E_{2T}^{'\nu} $ are shown against $x$ and $\Dp^2$. The left column  presents the plots for $u$ quark ((a), (c), (e) and (g)), and the right column for $d$ quark ((b), (d), (f) and (h)).}
\end{figure*}

\noindent
Twist-$3$ chiral-even GPDs have been analyzed using both $2$-D and $3$-D plots. The $2$-D plots illustrate the variation of GPDs with respect to one variable while keeping the other variable fixed. In contrast, the $3$-D plots depict simultaneous changes in GPDs concerning the variables $x$ and $\Delta_{\perp}$. The twist-$3$ chiral-even GPD $x \tilde{E}_{2T}^{\nu}$, concerning the matrix structure $\gamma^j$, has been plotted against $x$ and $\Delta_{\perp}$ for the active $u$ and $d$ quarks in Fig. (\ref{fig3dvxd}). The peaks of $x \tilde{E}_{2T}^{\nu}$ for active $u$ and $d$ quarks were found to exist in the low momentum transfer region, due to the occurrence of $\Delta_{\perp}$ in the denominator of Eqs. \eqref{xE2TBs} and \eqref{xE2TBv}. This feature of the GPD matches the findings in Ref. \cite{Zhang:2023xfe}. The maxima of the plots for active $u$ and $d$ quarks were found around $x=0.3$, suggesting the equal distribution of longitudinal momentum fraction $x$ among each of the three valence quarks of the proton. This GPD comprises only $S$-wave ($L_z = 0$). It was observed that the $T_{11}$ term corresponds to the parallel spins of active quark, and the parent proton contributes negatively, while for the anti-parallel alignment, the $T_{22}$ term contributes both negatively and positively to the distribution.
\par 
Now we refer to the GPDs connected with the matrix structure $\gamma^j \gamma_5$, which are $x H_{2T}^{'\nu}$, $x \tilde{H}_{2T}^{'\nu}$, and $x E_{2T}^{'\nu}$. These GPDs have been plotted in Fig. (\ref{fig3dvxd}) for both the possible flavors of struck quark. Observations indicate that the GPD $x H_{2T}^{'\nu}$ peaks at lower values of $\Delta_{\perp}$, and when switching the active quark flavor from $u$ to $d$, the peak of the distribution shifts slightly towards higher $x$. This GPD contains $S$-wave states from the $T_{11}$ term, a $P$-wave state (with $L_z = 1$) from the $T_{12}$ term, and a $D$-wave state (with $L_z = 2$) from the $T_{22}$ term. From Eqs. \eqref{xH2TPs} and \eqref{xH2TPv}, it is evident that the $S$-wave and $P$-wave contribute positively, while the $D$-wave contributes negatively to the distribution.
The GPD $x \tilde{H}_{2T}^{'\nu}$ is plotted in Figs. \ref{fig3dvxd} (e) and \ref{fig3dvxd} (f) for the $u$ and $d$ quarks, respectively. Similar to $x H_{2T}^{'\nu}$, this GPD also shows maxima at low values of $x$ and $\Delta_{\perp}$. Notably, the sign of the amplitude for the active $u$ quark distribution is positive, whereas for the active $d$ quark distribution, it is negative. From Eqs. \eqref{xH2TBPs} and \eqref{xH2TBPv}, it can be seen that $x \tilde{H}_{2T}^{'\nu}$ consists of a $P$-wave state from the $T_{12}$ term and a $D$-wave state from the $T_{22}$ term, both of which contribute oppositely to the GPD. Finally, the GPD $x E_{2T}^{'\nu}$ is represented in Figs. \ref{fig3dvxd} (g) and \ref{fig3dvxd} (h) for both active quark flavors $\nu$. The plots reveal significant similarities between $x \tilde{E}_{2T}^{\nu}$ and $x \tilde{H}_{2T}^{'\nu}$. From the overlap form of this GPD, it is observed that this GPD consists of other axial-vector GPDs as well. Apart from the $S$-wave, which only makes a negative contribution to the GPD, the $P$-wave and $D$-waves make both positive and negative contributions.
%
\par
\begin{figure*}
	\centering
	\begin{minipage}[c]{0.98\textwidth}
		(a)\includegraphics[width=0.44\textwidth]{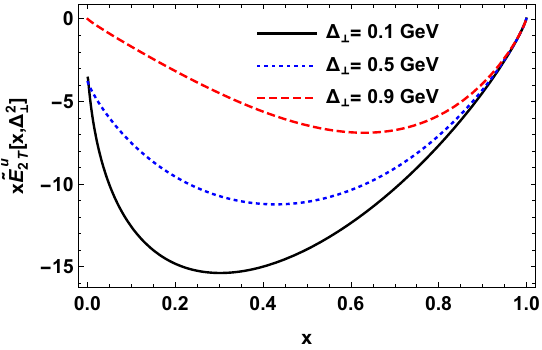}
		\hspace{0.05cm}
		(b)\includegraphics[width=0.44\textwidth]{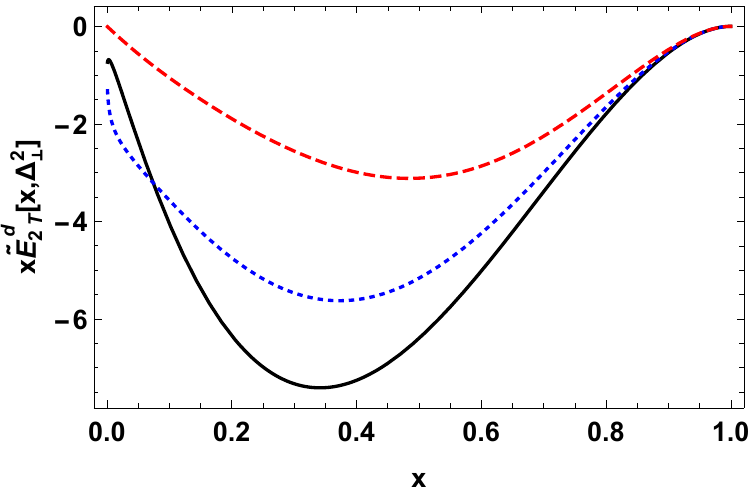}\\
		(c)\includegraphics[width=0.44\textwidth]{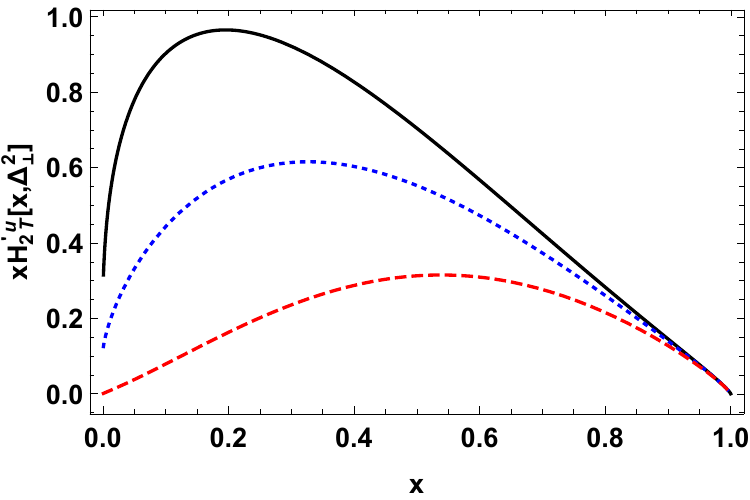}
		\hspace{0.05cm}
		(d)\includegraphics[width=0.44\textwidth]{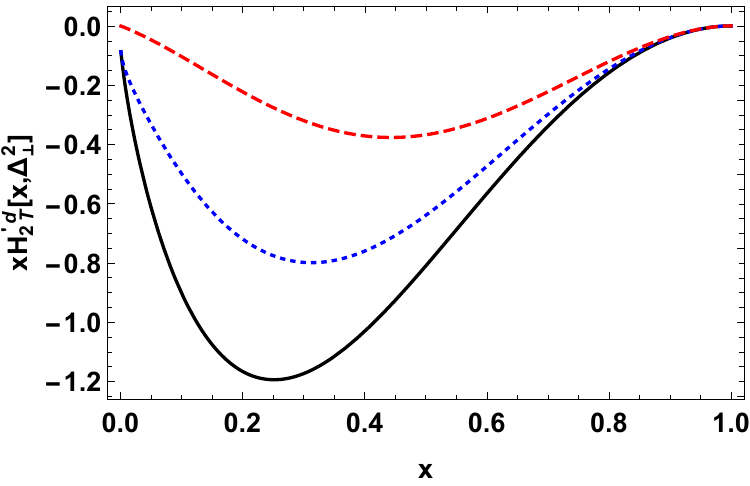}\\
		(e)\includegraphics[width=0.44\textwidth]{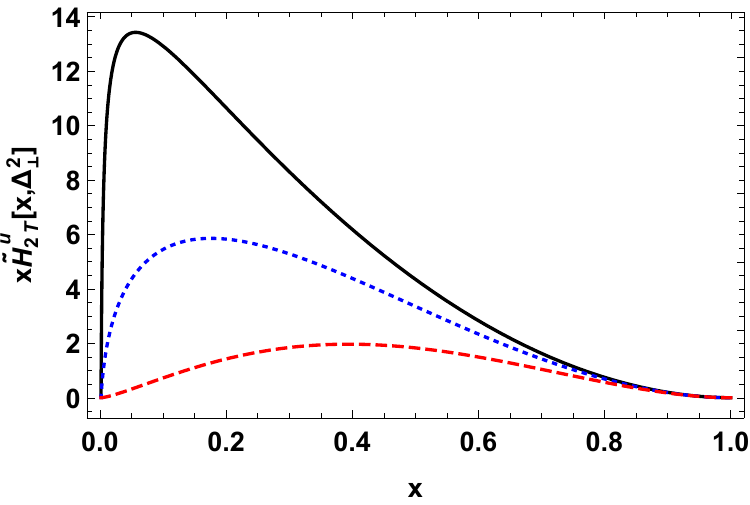}
		\hspace{0.05cm}
		(f)\includegraphics[width=0.44\textwidth]{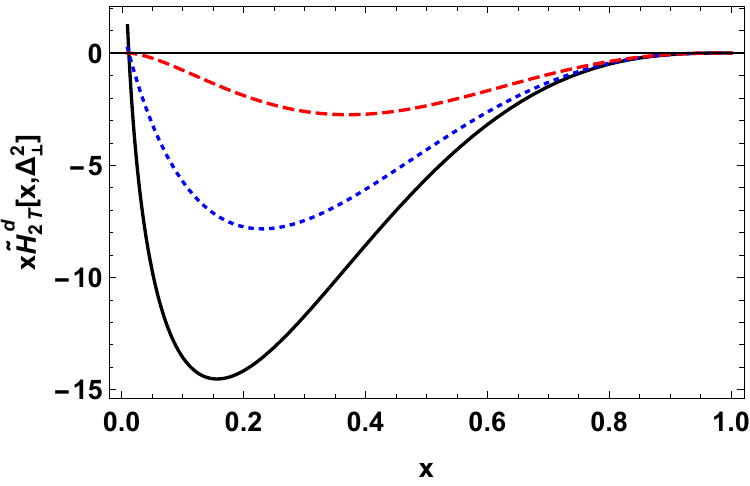}\\
		(g)\includegraphics[width=0.44\textwidth]{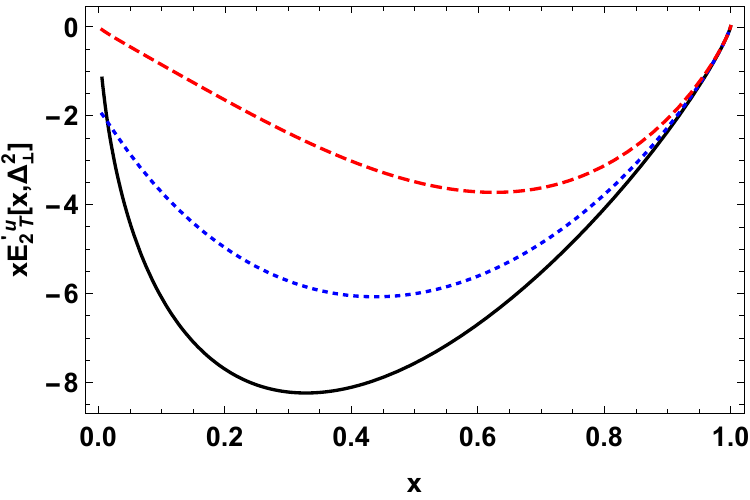}
		\hspace{0.05cm}
		(h)\includegraphics[width=0.44\textwidth]{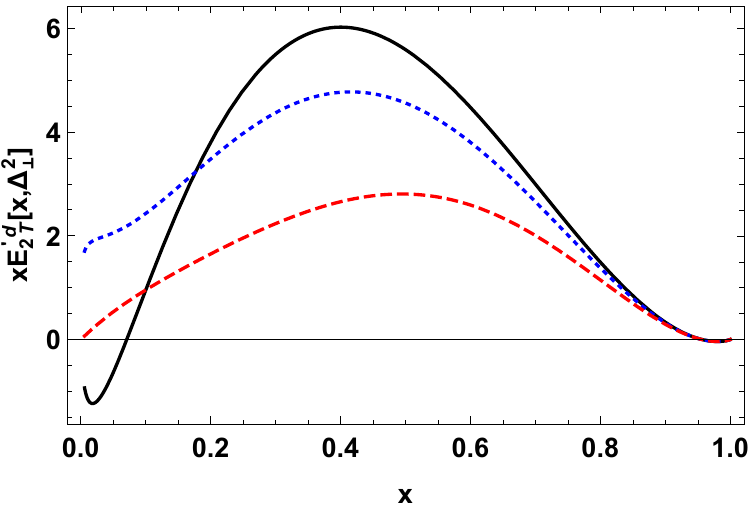}
		\\
	\end{minipage}
	\caption{\label{fig2dvx} The chiral-even twist-$3$ GPDs 		
		 $x \tilde{E}_{2T}^{\nu} $, $ x H_{2T}^{'\nu} $, $ x \tilde{H}_{2T}^{'\nu} $, and $ x E_{2T}^{'\nu} $ 
			plotted with respect to $x$ at various fixed values of $\Dp$. The left ((a), (c), (e) and (g)) and right ((b), (d), (f) and (h)) column correspond to $u$ and $d$ quarks sequentially.
	}
\end{figure*}
\begin{figure*}
	\centering
	\begin{minipage}[c]{0.98\textwidth}
		(a)\includegraphics[width=0.44\textwidth]{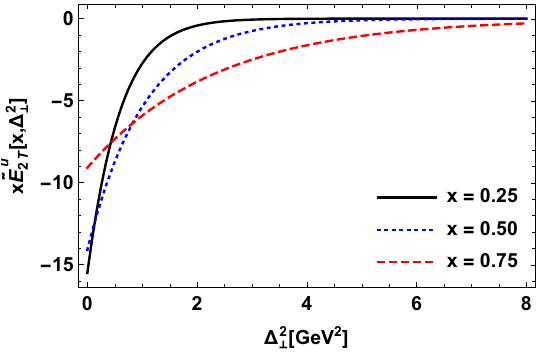}
		\hspace{0.05cm}
		(b)\includegraphics[width=0.44\textwidth]{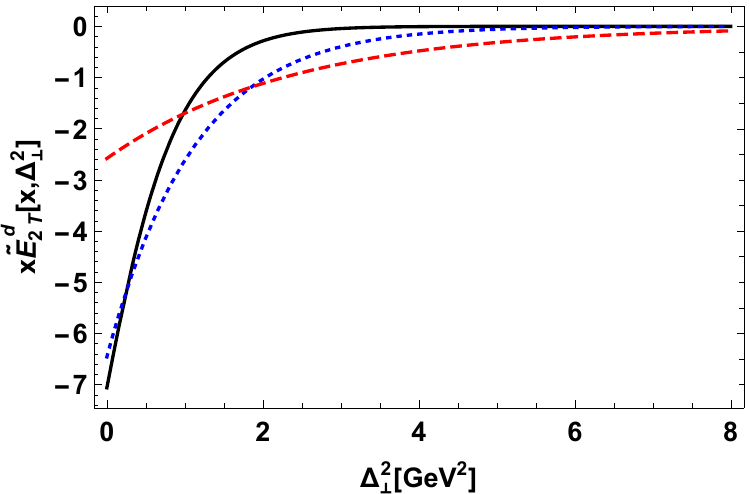}\\
		(c)\includegraphics[width=0.44\textwidth]{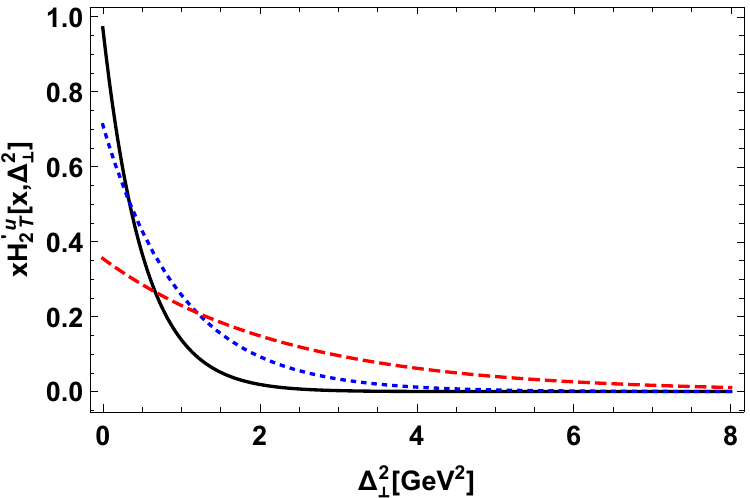}
		\hspace{0.05cm}
		(d)\includegraphics[width=0.44\textwidth]{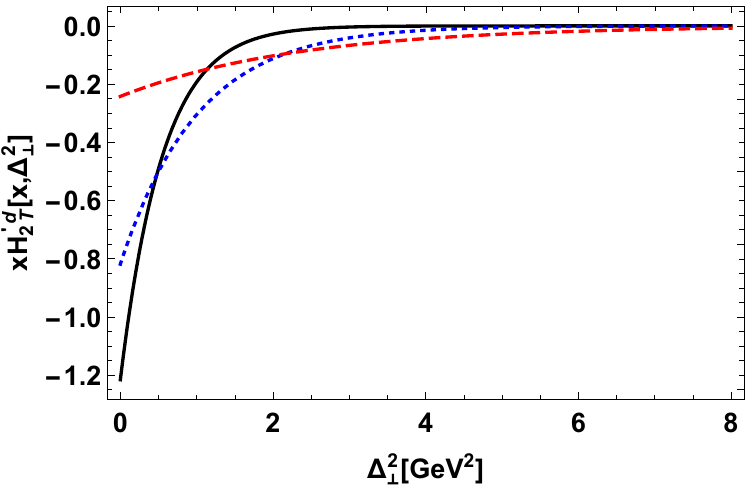}\\
		(e)\includegraphics[width=0.44\textwidth]{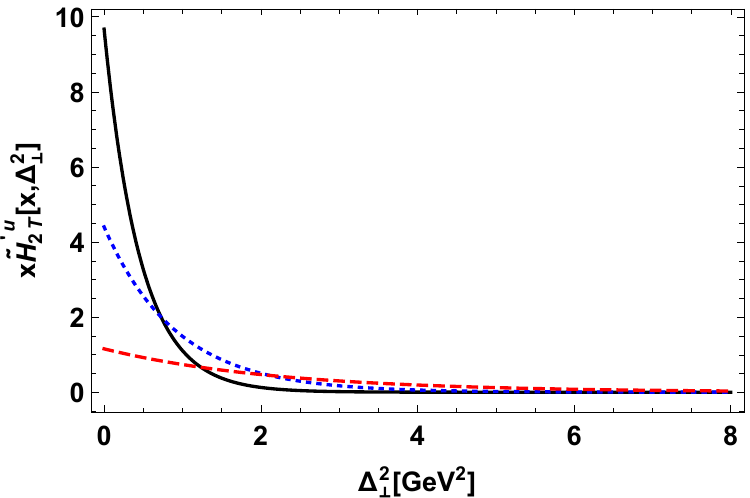}
		\hspace{0.05cm}
		(f)\includegraphics[width=0.44\textwidth]{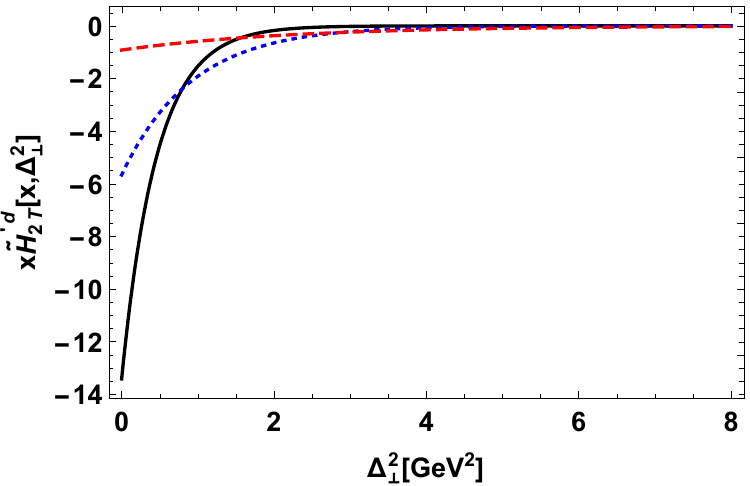}\\
		(g)\includegraphics[width=0.44\textwidth]{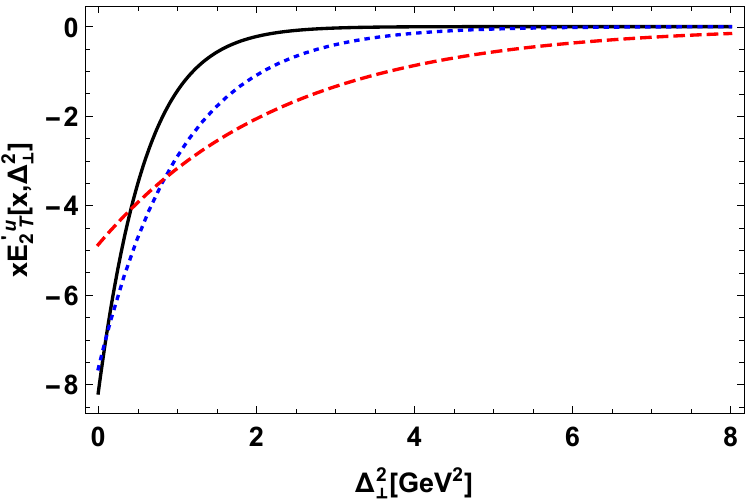}
		\hspace{0.05cm}
		(h)\includegraphics[width=0.44\textwidth]{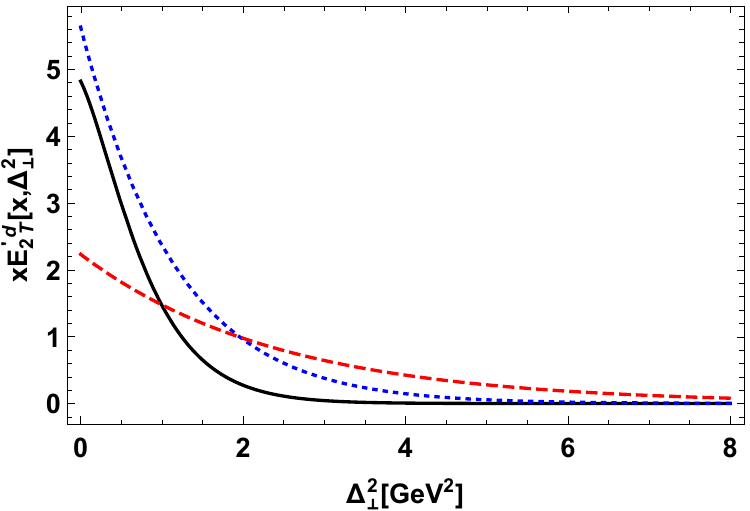}
		\\
	\end{minipage}
	\caption{\label{fig2dvd} The chiral-even twist-$3$ GPDs 		
		$x \tilde{E}_{2T}^{\nu} $, $ x H_{2T}^{'\nu} $, $ x \tilde{H}_{2T}^{'\nu} $, and $ x E_{2T}^{'\nu} $ 
			plotted with respect to $\Dp^2$ at various fixed values of $x$. The left ((a), (c), (e) and (g)) and right ((b), (d), (f) and (h))  column correspond to $u$ and $d$ quarks sequentially.
	}
\end{figure*}

To gain a better understanding of twist-$3$ chiral-even GPDs $x \tilde{E}_{2T}^{\nu}$, $x H_{2T}^{'\nu}$, $x \tilde{H}_{2T}^{'\nu}$, and $x E_{2T}^{'\nu}$, we have investigated their behavior in relation to the longitudinal momentum fraction $x$ at various fixed transverse momentum transfer values $\Delta_{\perp}$ for both active quark flavors ($\nu = u, d$). The key observations from Fig. (\ref{fig2dvx}) include that $x \tilde{E}_{2T}^{\nu}$ increases as the longitudinal momentum fraction $x$ increases, and after attaining a maximum value, it decreases for higher values of $x$ at different values of $\Delta_{\perp}$. The maxima of each GPD shifts towards higher $x$ as the transverse momentum transfer $\Delta_{\perp}$ is increased. In all GPDs, at very high values of $x$, the curves corresponding to the different values of $\Delta_{\perp}$ merge with each other. This suggests that at very high $x$, the momentum transfer $\Delta_{\perp}$ becomes ineffective for both $u$ and $d$ active quark flavors.
We also plot the GPDs in relation to $\Delta_{\perp}$ for different values of $x$ for both active quark flavors $u$ and $d$. From Fig. (\ref{fig2dvd}), it is apparent that all the GPDs follow a similar trend. As $x$ increases to high values, the peak of the distributions becomes broader, suggesting that the contributions to higher $\Delta_{\perp}$ come from high values of $x$. Conversely, for low values of $x$, such as $0.25$, the plots become sharp, and negligible contributions are seen for high values of $\Delta_{\perp}$.


\subsection{{Impact parameter dependent parton distributions}\label{ssipdpdfs}}
\begin{figure*}
	\centering
	\begin{minipage}[c]{0.98\textwidth}
		(a)\includegraphics[width=0.44\textwidth]{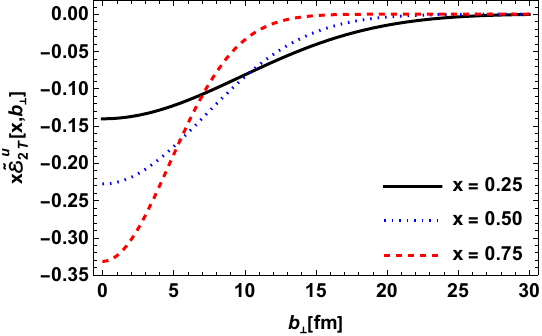}
		\hspace{0.05cm}
		(b)\includegraphics[width=0.44\textwidth]{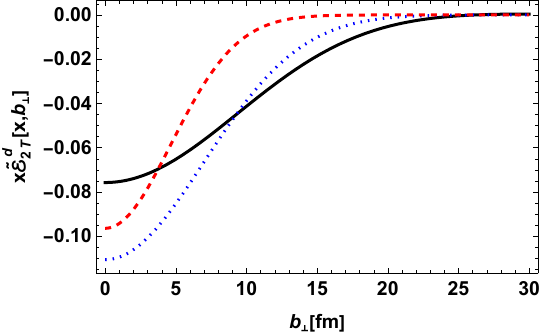}\\
		(c)\includegraphics[width=0.44\textwidth]{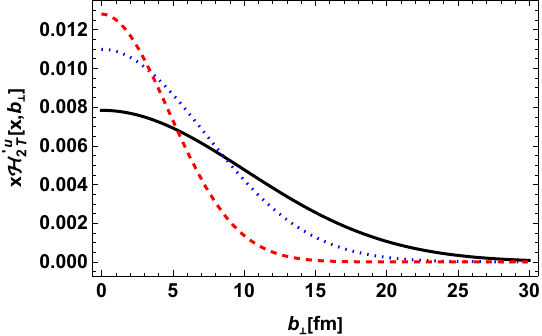}
		\hspace{0.05cm}
		(d)\includegraphics[width=0.44\textwidth]{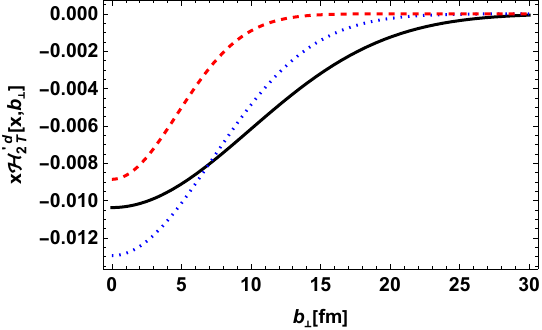}\\
		(e)\includegraphics[width=0.44\textwidth]{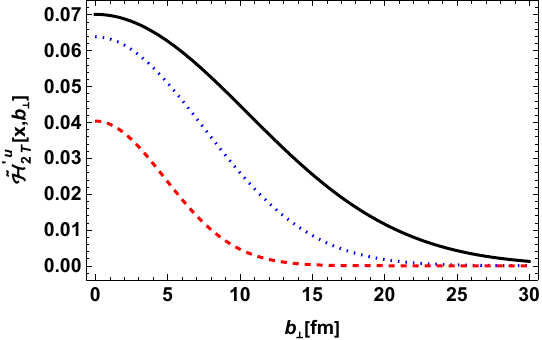}
		\hspace{0.05cm}
		(f)\includegraphics[width=0.44\textwidth]{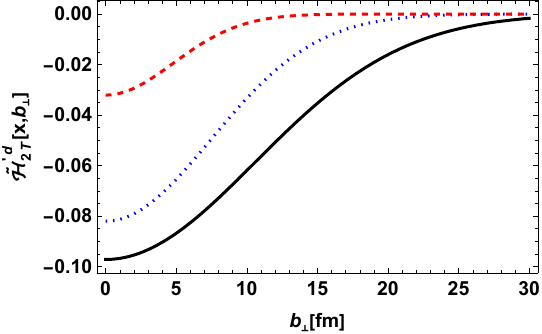}\\
		(g)\includegraphics[width=0.44\textwidth]{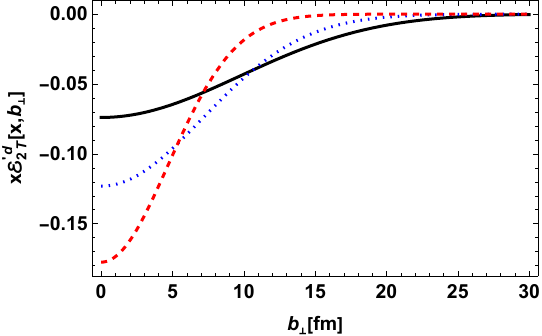}
		\hspace{0.05cm}
		(h)\includegraphics[width=0.44\textwidth]{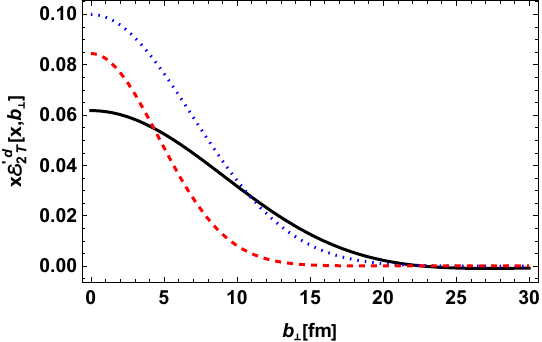}
		\\
	\end{minipage}
	\caption{\label{figFT2dvb} The twist-$3$ IPDPDFs 	
		$x 	\mathcal{\tilde{E}}_{2T}^{\nu}$, $	x \mathcal{ H}_{2T}^{'\nu}$, $	x \mathcal{ \tilde{H}}_{'2T}^{\nu}$, and $x	\mathcal{ E}_{2T}^{'\nu}$
		plotted with respect to $\bf{b_\perp}$ at various fixed values of $x$. The left ((a), (c), (e) and (g)) and right ((b), (d), (f) and (h))  column correspond to $u$ and $d$ quarks sequentially.
	}
\end{figure*}

\begin{figure*}
	\centering
	\begin{minipage}[c]{0.98\textwidth}
		(a)\includegraphics[width=0.39\textwidth]{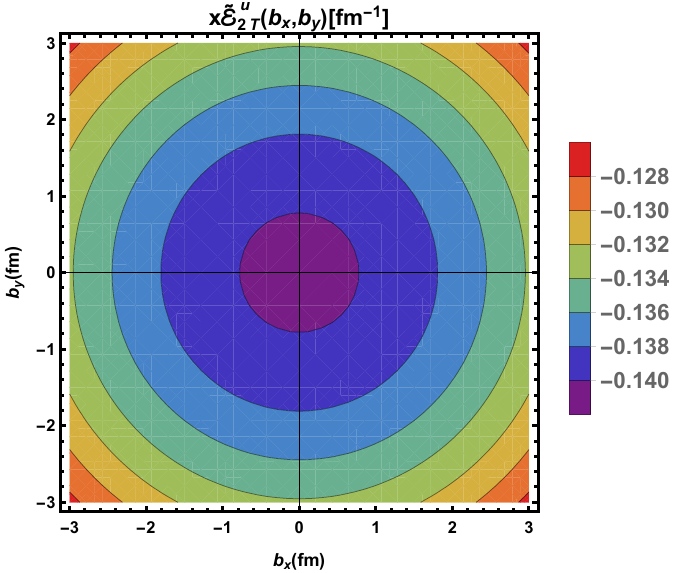}
		(b)\includegraphics[width=0.39\textwidth]{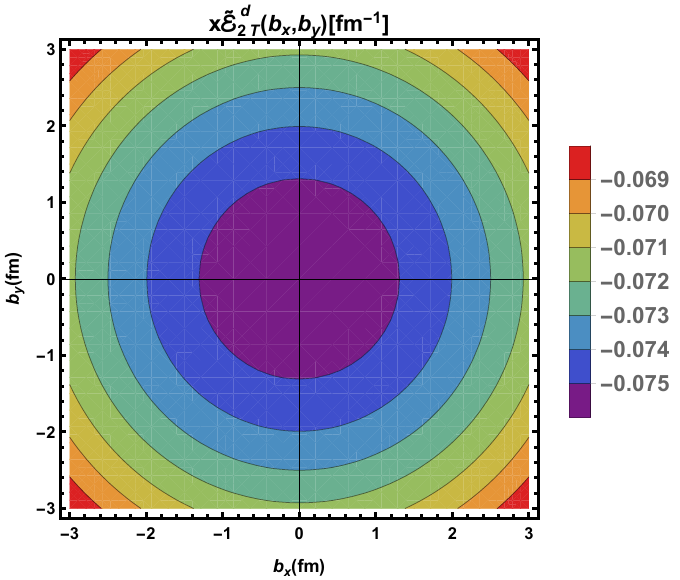}\\
		(c)\includegraphics[width=0.39\textwidth]{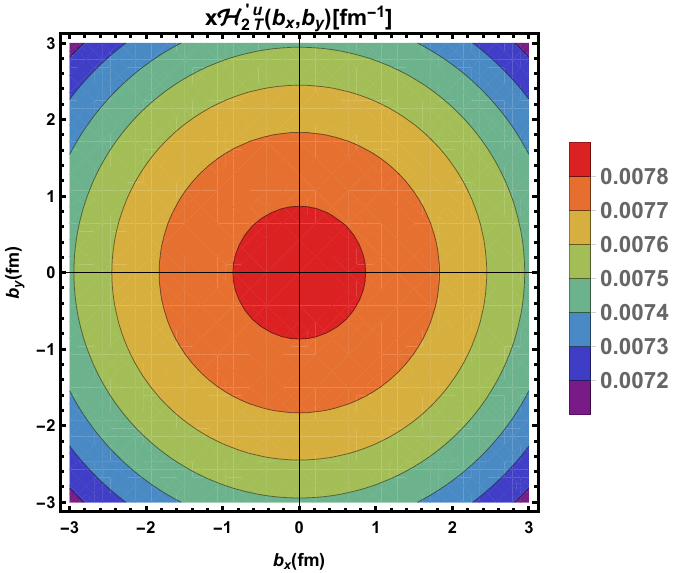}
		\hspace{0.05cm}
		(d)\includegraphics[width=0.39\textwidth]{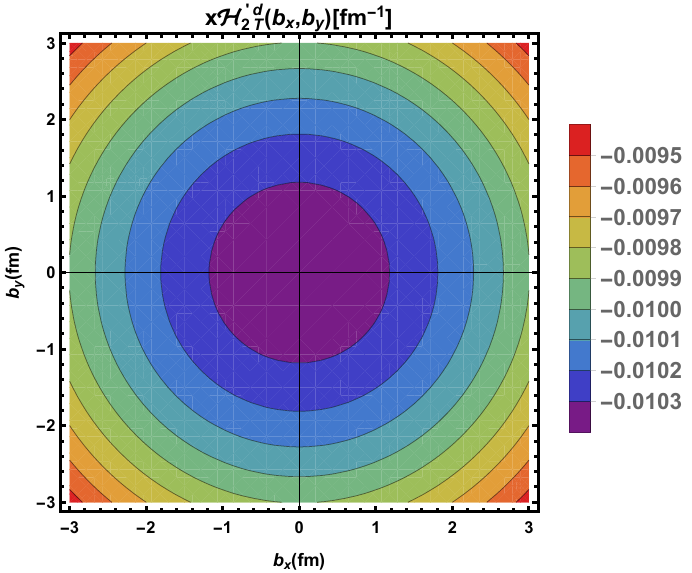}\\
		(e)\includegraphics[width=0.39\textwidth]{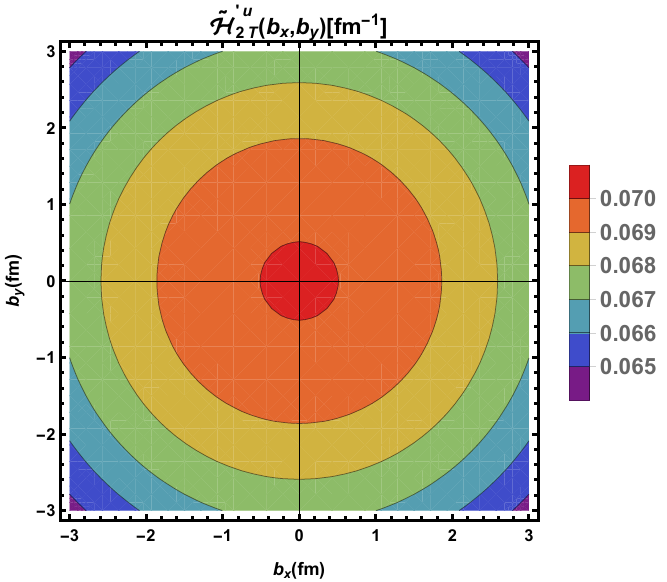}
		(f)\includegraphics[width=0.39\textwidth]{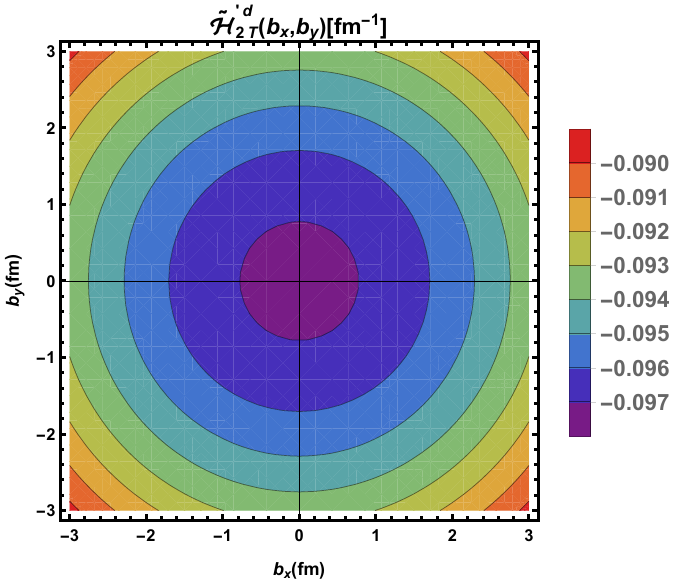}\\
		(g)\includegraphics[width=0.39\textwidth]{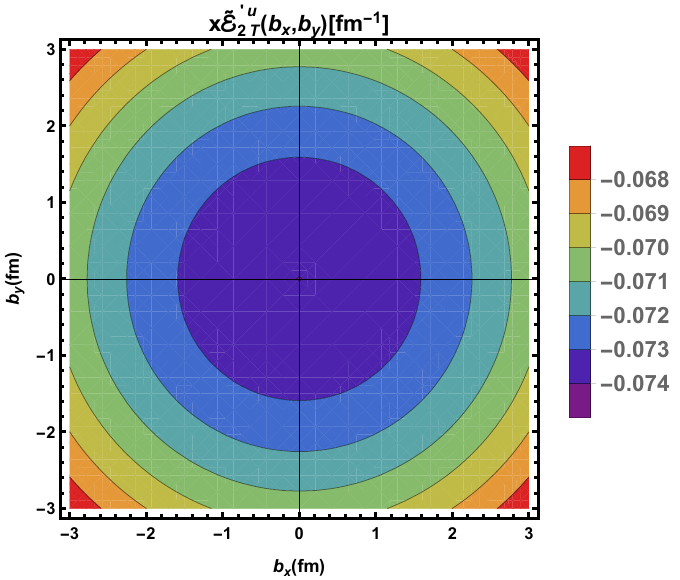}
		(h)\includegraphics[width=0.39\textwidth]{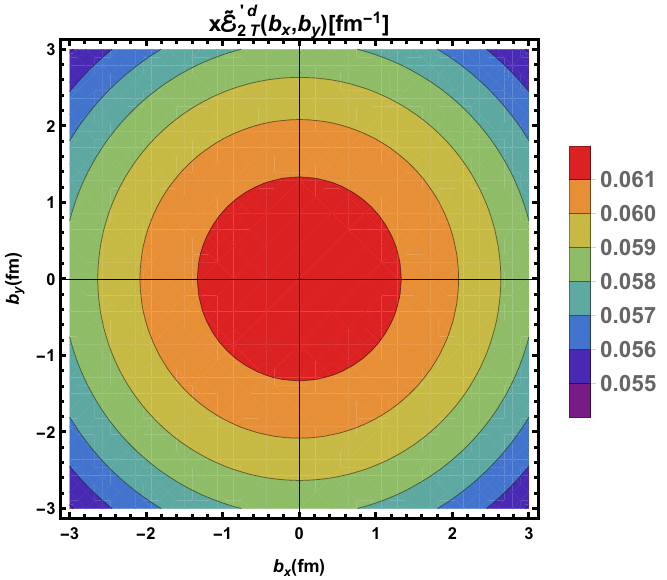}
	\end{minipage}
	\caption{\label{figFTCP} The twist-$3$ IPDPDFs 	
		$x 	\mathcal{\tilde{E}}_{2T}^{\nu}$, $	x \mathcal{ H}_{2T}^{'\nu}$, $	x \mathcal{ \tilde{H}}_{'2T}^{\nu}$, and $x	\mathcal{ E}_{2T}^{'\nu}$
		plotted with respect to $b_x$ and $b_y$  at various fixed values of $x$. The left ((a), (c), (e) and (g)) and right ((b), (d), (f) and (h))  column correspond to $u$ and $d$ quarks sequentially.
	}
\end{figure*}

\noindent
A distinct viewpoint on the spatial parton distribution within hadrons is offered by the impact parameter space GPDs, also known as IPDPDFs \cite{Burkardt:2000za}. These distributions offer an additional viewpoint on the spatial distribution of partons within hadrons by describing the probability amplitude of locating a parton at a certain transverse distance $\bfb$ from the hadron center. The Fourier transform is carried out in $\Dp$ to produce IPDPDFs as \cite{Burkardt:2002hr}
\begin{equation}
	\mathcal{X^{\nu}}(x,\bfb)=\frac{1}{(2\pi)^2} \int d^{2}\Dp e^{-i b_{\perp}\cdot\Dp}X^{\nu}(x,\Dp^2). \\
\end{equation}
Here, $X^{\nu}(x,\Dp^2)$ and $\mathcal{X^{\nu}}(x,\bfb)$ denotes the corresponding GPD and IPDPDF sequentially. We have plotted twist-$3$ IPDPDFs ($x \mathcal{\tilde{E}}_{2T}^{\nu}$, $x\mathcal{ H}_{2T}^{'\nu}$,$	x \mathcal{ \tilde{H}}_{'2T}^{\nu}$, and $x\mathcal{ E}_{2T}^{'\nu}$) in Fig.(\ref{figFT2dvb}) for both the possibilities of active quark flavor being  $u$ or $d$. IPDPDF $x \mathcal{\tilde{E}}_{2T}^{\nu}$  has been plotted in Figs. \ref{figFT2dvb} (a) and \ref{figFT2dvb} (b), which shows that $x \mathcal{\tilde{E}}_{2T}^{\nu}$ has high probability of being concentrated towards center of momentum (COM) line and as the longitudinal momentum fraction $x$ is increased the peak of distribution for active $u$ becomes more negative and approaches to zero for higher values of $b_{\perp}$ while for active $d$ quark distribution, the maximum peak is observed for longitudinal momentum fraction being about $0.5$. Now, while observing plots in Figs. \ref{figFT2dvb} (c) and \ref{figFT2dvb} (d), it can be realized that the for the polarization configuration corresponding to IPDPD $x \mathcal{ H}_{2T}^{'\nu}$ that the on the change of active quark flavor the sign of the amplitude of the distribution reverses. For active $d$ quark, the distribution corresponding to $x \mathcal{ H}_{2T}^{'\nu}$ is similar to that of $x 	\mathcal{\tilde{E}}_{2T}^{\nu}$, keeping magnitude aside. In Figs. \ref{figFT2dvb} (e) and \ref{figFT2dvb} (f), IPDPDF $x \mathcal{ \tilde{H}}_{'2T}^{\nu}$ is plotted for each active quark flavor.While considering the behavior of distribution along the longitudinal momentum fraction $x$, plots of $x \mathcal{ \tilde{H}}_{'2T}^{\nu}$ show the most distinct trend than the remaining distributions in a sense that the lines corresponding to different $x$ are not tangled. Furthermore, it has been noted that when $x$ decreases, the distribution's magnitude grows for both active quark flavors $(\nu = u,d)$. For the IPDPDF $x \mathcal{E}_{2T}^{'\nu}$, the plots corresponding to this distribution are found to be remarkably similar in trend to those of $x \mathcal{\tilde{E}}_{2T}^{\nu}$, although they differ in amplitude. Another distinction is that the plots for $x \mathcal{\tilde{E}}_{2T}^{\nu}$ exhibit the same polarity for different flavors of the active quark, whereas $x \mathcal{E}_{2T}^{'\nu}$ does not.
To provide a clearer understanding of the possibility of the active quark being near the center of mass (COM), we have depicted the distributions using contour plots in Fig. \ref{figFTCP}, where GPDs are shown as functions of the transverse coordinates $ b_x $ and $ b_y $. These plots allow us to visualize the spatial distribution of quarks within the hadron, giving us insights into how the quark density varies in the transverse plane.

\subsection{{Form Factors}\label{ssffs}}
\begin{figure*}
	\centering
	\begin{minipage}[c]{0.98\textwidth}
		(a)\includegraphics[width=0.44\textwidth]{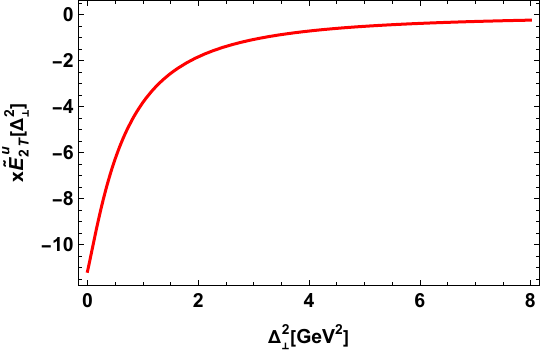}
		\hspace{0.05cm}
		(b)\includegraphics[width=0.44\textwidth]{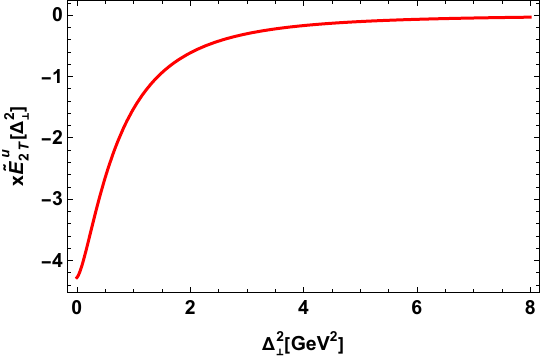}\\
		(c)\includegraphics[width=0.44\textwidth]{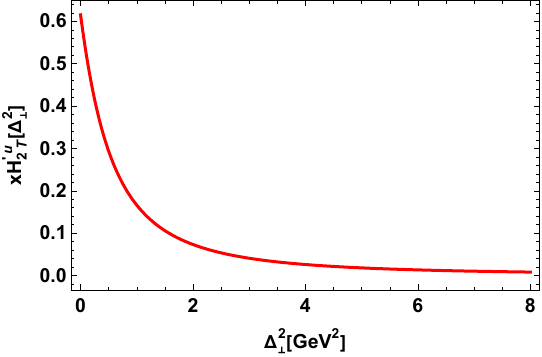}
		\hspace{0.05cm}
		(d)\includegraphics[width=0.44\textwidth]{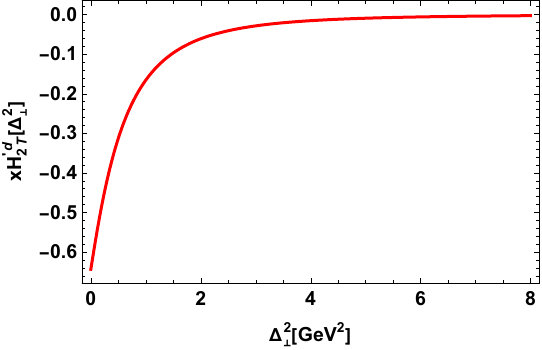}\\
		(e)\includegraphics[width=0.44\textwidth]{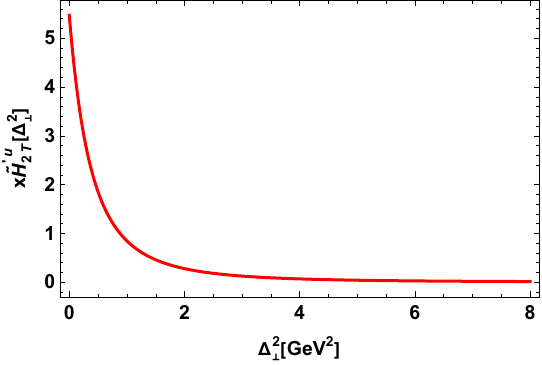}
		\hspace{0.05cm}
		(f)\includegraphics[width=0.44\textwidth]{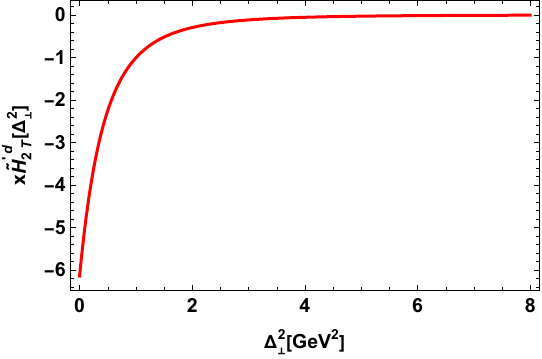}\\
		(g)\includegraphics[width=0.44\textwidth]{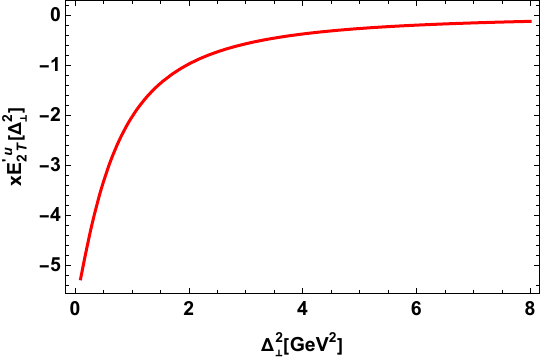}
		\hspace{0.05cm}
		(h)\includegraphics[width=0.44\textwidth]{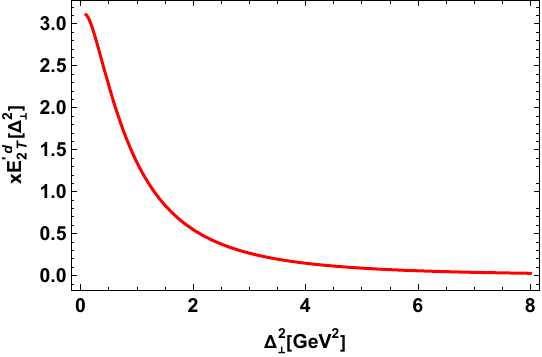}
		\\
	\end{minipage}
	\caption{\label{fig2dffvd} The twist-$3$ FFs 		
		$x \tilde{E}_{2T}^{\nu} $, $ x H_{2T}^{'\nu} $, $ x \tilde{H}_{2T}^{'\nu} $, and $ x E_{2T}^{'\nu}$ 
	plotted with respect to $\Dp^2$. The left ((a), (c), (e) and (g)) and right ((b), (d), (f) and (h))  column correspond to $u$ and $d$ quarks sequentially.
	}
\end{figure*}

\noindent
In Fig.  (\ref{fig2dffvd}), We have plotted the twist-$3$ FFs ( $x \tilde{E}_{2T}^{\nu}$, $x H_{2T}^{'\nu}$, $x \tilde{H}_{2T}^{'\nu}$, and $x E_{2T}^{'\nu}$) obtained   by integrating twist-$3$ chiral-even GPDs over the longitudnal momentum fraction $x$. At first glance, the magnitude of all the FF plots appears to exponentially approach to zero when the transverse momentum transfer to the proton square $(\Delta_{\perp}^2)$ reaches higher values. This exponential trend underscores the strong suppression of quark distributions at higher transverse momentum transfers, highlighting the intricate spatial configuration of the proton's internal structure.

\section{{Conclusion}\label{seccon}}
\noindent
In our study of twist-$3$ chiral-even GPDs within the LFQDM, we have derived the expressions for the concerned GPDs by solving the appropriate parametrization equations. Using the LFWFs for both scalar and vector diquark configurations, we have obtained the GPDs for both possible cases of diquarks. Explicit equations were derived for the scenarios where the active quark flavor is either $u$ or $d$.
We have utilized the relationship between the GTMD correlator and the GPD correlator to establish the connection between twist-$3$ chiral-even GPDs and twist-$3$ GTMDs. Most of these relations are consistent with those presented in Ref. \cite{Meissner:2009ww}. Some of our results are also in agreement with findings from the BLFQ approach \cite{Zhang:2023xfe} and the Lattice QCD \cite{Bhattacharya:2023jsc}.
We have done a detailed discussion on twist-$3$ GPDs, IPDPDFs, and FFs by employing $2$-D and $3$-D plots to illustrate the variation of these expressions with respect to the kinematic variables, providing a comprehensive understanding of their behavior.
\par
The discovery of exclusive reactions involving GPDs is about to enter a new era with the arrival of the Electron-Ion Collider (EIC) at Brookhaven National Laboratory (BNL). With its cutting-edge capabilities, the EIC is poised to deliver unprecedented precision in experimental data, enabling a more thorough examination of the strong force and the refinement of existing GPD models. By enhancing Lattice QCD calculations and developing sophisticated phenomenological models, we can better interpret the plethora of data from such advanced experiments. GPDs have already broadened our understanding of proton structure, and the precise data provided by these advanced tools will enable us to gain deeper insights into the complexities of the strong force and assess the limitations of current GPD models. Future theoretical works on GPDs, such as calculating non-zero skewness distributions with the inclusion of higher Fock-states across different proton polarizations, will be crucial to support these experimental developments. In the future, it would be fascinating to explore model-independent calculations of these higher twist distributions, especially in the low-$x$ regime.
%

\section*{ACKNOWLEDGMENTS}

H.D. would like to thank  the Science and Engineering Research Board, Anusandhan-National Research Foundation, Government of India under the scheme SERB-POWER Fellowship (Ref No. SPF/2023/000116) for financial support.

\end{document}